\documentclass[%
 reprint,
 amsmath,amssymb,
 aps,
]{revtex4-2}
\usepackage{mathptmx}
\usepackage{etoolbox}
\usepackage{xcolor}
\usepackage{hyperref}
\usepackage{graphicx}% Include figure files
\usepackage{dcolumn}% Align table columns on decimal point
\usepackage{bm}% bold math
\begin{document}

\preprint{APS/123-QED}

\title{Two-gap superconductivity in a Janus MoSeLi monolayer.}% Force line breaks with \\
%\thanks{A footnote to the article title}%

\author{Jakkapat Seeyangnok}
 \email{jakkapatjtp@gmail.com}
\affiliation{%
 Department of Physics, Faculty of Science, Chulalongkorn University, 254 Phaya Thai Rd, Bangkok, 10330, Thailand.
}

\author{Udomsilp Pinsook}%
 \email{Udomsilp.P@Chula.ac.th.}
\affiliation{%
 Department of Physics, Faculty of Science, Chulalongkorn University, 254 Phaya Thai Rd, Bangkok, 10330, Thailand.
}

\author{Greame John Ackland}
 \email{gjackland@ed.ac.uk}
\affiliation{%
 Centre for Science at Extreme Conditions, School of Physics and Astronomy, University of Edinburgh, Edinburgh, EH9 3FD, Scotland, United Kingdom.
}%

%\collaboration{CLEO Collaboration}%\noaffiliation

\date{\today}% It is always \today, today,
             %  but any date may be explicitly specified

\begin{abstract}
Two-dimensional (2D) lithium-decorated materials have emerged as a significant area of study since the prediction of superconductivity in lithium-decorated graphene at temperatures around 8.1 K (G. Profeta et al., Nature Physics 8, 131–134) , with experimental evidence observed at $T_c$ = 5.9 K (B. M. Ludbrook et al., PNAS 112 (38) 11795-11799). Following earlier studies, this paper focuses on the hexagonal Janus MoSeLi monolayer as a promising candidate for Li-decorated 2D materials. Our research reveals that lithium substitution on a selenium layer of MoSe$_2$ can produce a hexagonal Janus MoSeLi monolayer, which exhibits metallic behavior with potential phonon-mediated superconductivity with a critical temperature (\(T_c\)) of 4.5 K. Additionally, by solving the anisotropic gap equations derived from the Migdal-Eliashberg theory, we found that the Janus MoSeLi monolayer exhibits two-gap superconductivity. This finding underscores the potential of hexagonal Janus MoSeLi as a significant Li-decorated 2D material for exploring two-dimensional superconductivity and sets the stage for further investigations into new families of Janus transition metal chalcogenides.
\end{abstract}

%\keywords{Suggested keywords}%Use showkeys class option if keyword
                              %display desired
\maketitle

%\tableofcontents
\section{Introduction}
Two-dimensional (2D) lithium-decorated materials have emerged as a significant area of study since the prediction of superconductivity in lithium-decorated graphene\cite{profeta2012phonon}. The decoration of lithium in graphene improves its electronic properties and has been shown to induce superconductivity at temperatures around 8.1 K\cite{profeta2012phonon}, with experimental observation at 5.9 K\cite{ludbrook2015evidence} using angle-resolved photoelectron spectroscopy (ARPES). ARPES experiment allows for the investigation of various physical properties of quasielectrons resulting from the electron-phonon interaction. Following this, lithium-decorated materials have been extensively studied. Examples include lithium-decorated 2D orthorhombic (o)-B$_2$X$_2$ monolayers \cite{benaddi2024lithium}, lithium-decorated SiB monolayer \cite{jiang2023lithium}, lithium-decorated 2D Irida-graphene \cite{zhang2024li}, lithium-decorated phosphorene \cite{haldar2017first} and lithium-decorated $\Psi$-graphene \cite{dewangan2023lithium}. These advancements make lithium-decorated 2D materials promising candidates for advanced electronic applications and superconducting devices.

2D Janus transition-metal dichalcogenides (2D-JTMDs) exhibit an out-of-plane asymmetric layered structure, with different chalcogenide atoms on each side. This asymmetry imparts 2D-JTMDs with unique and tunable electronic, optical, and mechanical properties \cite{tang20222d,zhang2022janus,angeli2022twistronics,he2018two,yeh2020computational,yin2021recent}. Although 2D-JTMDs do not occur naturally, they have been successfully synthesized, beginning with Janus graphene in 2013 \cite{zhang2013janus}. Since then, various 2D-JTMDs have been developed, such as MoSSe \cite{trivedi2020room,lu2017janus}, WSSe \cite{trivedi2020room}, and PtSSe \cite{sant2020synthesis}. These materials are typically fabricated using selective epitaxy atomic replacement (SEAR) processes with hydrogen (H$_2$) plasma \cite{trivedi2020room,tang20222d}. Recently, a Janus 2H-MoSH monolayer was synthesized using the SEAR method by substituting the top S atoms with hydrogen atoms \cite{lu2017janus}. Theoretically, Janus MoSH monolayers have been predicted to exhibit superconductivity with $T_c = 26.81$ K for the 2H phase \cite{liu2022two,ku2023ab}. Furthermore, the 2H and 1T phases of WSH and WSeH phase has been suggested to be dynamically stable with a $T_c$ around 12 K \cite{seeyangnok2024superconductivity,seeyangnok2024superconductivitywseh}. These phase stability and superconductivity have been confirmed by subsequent independent studies \cite{gan2024hydrogenation,fu2024superconductivity}. Recently, possible superconductivity has been reported in hexagonal TiSH, TiSeH, TiTeH, and ZrTeH monolayers \cite{ul2024superconductivity,li2024machine}. Consequently, the study of elemental substitutions in Janus transition-metal dichalcogenides has been extensively pursued. Additionally, substituting lithium is a novel technique to modify the surface properties of 2D materials, as it might enhance electronic and physical characteristics of those 2D materials \cite{singh2021review,chen2021interstitial,xie2024strong}.

In this paper, we aims to comprehensively study the 2H hexagonal Janus MoSeLi monolayer as a potential Li-decorated 2D material, specifically in the form of hexagonal Janus transition metal dichalcogenide monolayer. First, we discuss the electronics to understand the properties of electrons near the Fermi level, and examine the phonon properties to study the vibrational modes of collective phonons. Subsequently, we investigate superconductivity through electron-phonon interactions based on the Migdal-Eliashberg (ME) theory \cite{bardeen1957microscopic,frohlich1950theory,migdal1958interaction,eliashberg1960interactions,nambu1960quasi}. This framework allows us to explore the multigap superconductivity arising from the anisotropic electron-phonon interaction. The self-consistent calculations implemented QUANTUM ESPRESSO (QE) \cite{giannozzi2009quantum,giannozzi2017advanced} where we compute the electronics and phonon properties. We employed the electron-phonon Wannier-Fourier interpolation method \cite{giustino2017electron,giustino2007electron} within the EPW package \cite{noffsinger2010epw,ponce2016epw} to self-consistently solve anisotropic Migdal-Eliashberg equations. The calculation detail will be available in the supplementary.
\section{Methodology}
The computations were performed using density functional theory (DFT), as implemented in QUANTUM ESPRESSO (QE) \cite{giannozzi2009quantum,giannozzi2017advanced}. The crystal structures were created using VESTA \cite{momma2011vesta} within the trigonal space group \( P\overline{3}m1 \) (No.156). Optimization of the crystal structures was achieved using BFGS algorithm \cite{BFGS,liu1989limited}, with full relaxation to a force threshold of $10^{-5}$ eV/\AA. Optimized norm-conserving Vanderbilt pseudopotentials \cite{hamann2013optimized,schlipf2015optimization} with GGA-PBE \cite{perdew1996generalized} were employed for the exchange-correlation functional, with wavefunction and charge density cutoffs of 80 Ry and 320 Ry, respectively. For self-consistent Kohn-Sham states, a Monkhorst-Pack k-point grid \cite{monkhorst1976special} of $24 \times 24 \times 1$ with Methfessel-Paxton smearing of 0.02 Ry was used on the Fermi surface \cite{methfessel1989high}. The magnetic phases of MoSeLi were tested, and we found that the non-magnetic metallic phases are energetically favorable. The Fermi surface were visualized using XCRYSDEN \cite{kokalj2003computer}. Dynamical matrices were computed using density functional perturbation theory (DFPT) on $12 \times 12 \times 1$ q-point grids, for phonon self-consistent calculations. Then, we computed the phonon linewidth,
	\begin{equation}  \label{gammaphononlinewidths}
		\gamma_{\boldsymbol{q}\nu} = 2\pi\omega_{\boldsymbol{q}\nu}\sum_{nm}\sum_{\boldsymbol{k}}|g_{\boldsymbol{k}+\boldsymbol{q},\boldsymbol{k}}^{\boldsymbol{q}\nu,mn}|^{2}\delta(\epsilon_{\boldsymbol{k}+\boldsymbol{q},m}-\epsilon_{F})\delta(\epsilon_{\boldsymbol{k},n}-\epsilon_{F})
	\end{equation}
     and the electron-phonon coupling associated with phonon wavevector $\boldsymbol{q}$ and phonon modes of $\nu$, as
    \begin{equation}\label{eqn:lambda_qv}
        \lambda_{\boldsymbol{q}\nu} = \frac{\gamma_{\boldsymbol{q}\nu}}{\pi N(\epsilon_{\textbf{F}})\omega_{\boldsymbol{q}\nu}^2}.
    \end{equation}

Moreover, we can consider the electron self-energy due to electron-phonon coupling at the lowest order of the one-phonon bubble of the Feynman diagram. The real part of electron self-energy can be expressed as
    \begin{equation}\label{Real-part-equation}
        Re(\Sigma(k,u)) = -\int d\omega \alpha^{2}_{k} F(\omega)\log{\left|\frac{u+\omega}{u-\omega}\right|},
    \end{equation}
    where $\alpha^{2}_{k} F(\omega)=\int \frac{d^3 q}{(2\pi)^3} \delta(\omega-\omega(q))\delta(\xi_{k+q})|g_{k+q,k}|^2$ is averaged coupling function or Eliashberg spectral function. This electronic self energy reflects the cumulative effect integrated across the entire Brillouin zone (BZ).

    We employed the electron-phonon Wannier-Fourier interpolation method \cite{giustino2017electron,giustino2007electron} within the EPW package \cite{noffsinger2010epw,ponce2016epw} to accurately compute the superconducting properties, including the electron-phonon coupling constant ($\lambda$) and the Eliashberg spectral function ($\alpha^{2} F(\omega)$). This method also enabled the analysis of the anisotropic Migdal-Eliashberg theory \cite{margine2013anisotropic} through numerical solutions of the two nonlinear coupled anisotropic Migdal-Eliashberg equations,
    \begin{equation}
        Z_{nk}(i\omega_{j}) = 1+\frac{\pi T}{N({\varepsilon_{F})}\omega_{j}}\sum_{mk'j'}\frac{\omega_{j'}}{\sqrt{\omega^{2}_{j'}+\Delta^{2}_{mk'}(i\omega_{j'})}}, \\
    \end{equation}
    and
    \begin{eqnarray}
        Z_{nk}(i\omega_{j})\Delta_{mk}(i\omega_{j}) &= \frac{\pi T}{N(\varepsilon_{F})} \sum_{mk'j'} \frac{\Delta_{mk'}(i\omega_{j'})}{\sqrt{\omega^{2}_{j'}+\Delta^{2}_{mk'}(i\omega_{j'})}}\delta(\epsilon_{mk'}-\varepsilon_F)\nonumber \\     
    &\times[\lambda(nk,mk',\omega_j -\omega_{j'})-\mu^*].
    \end{eqnarray}
    
    We solved these equations self-consistently along the imaginary axis at the fermion Matsubara frequencies $\omega_j = (2j+1)\pi T$. For these computations, we employed the QUANTUM ESPRESSO package using self-consistent calculations as mentioned above. This was followed by Wannier-Fourier interpolation to k- and q-point grids of 120$\times$120$\times$1 and 60$\times$60$\times$1, respectively. The use of dense grids ensures the convergence of $\lambda$ values, as evidenced by the stability of the corresponding $\alpha^{2} F(\omega)$ and $\lambda(\omega)$ with increasing k- and q-point grid densities. The Fermi surface thickness was set to 0.25 eV, with the Matsubara frequency cutoff at 0.6 eV. The Dirac $\delta$ functions were broadened using a Gaussian function with widths of 0.0625 eV for electrons and 0.5 meV for phonons. Finally, the Morel-Anderson pseudopotential was set to $\mu^* = 0.1$ for practical purposes.

    The critial temperature of superconducting transition (T$_{C}$) was calculated using the semi-empirical Allen-Dynes formula \cite{allen1975transition} as
    \begin{equation}
    T_{c} = f_1 f_2 \frac{\omega_{\text{log}}}{1.20} \exp\left(-\frac{1.04(1+\lambda)}{\lambda-\mu^{*}(1+0.62\lambda)}\right)
\end{equation}
where the electron-phonon coupling ($\lambda$) can be computed from the Eliashberg spectral function ($\alpha^2 F(\omega)$) by using 
\begin{equation}
    \lambda = 2\int^{\omega}_{0} d\Omega \left(\frac{\alpha^{2}F(\Omega)}{\Omega}\right),
\end{equation}
and the logarithmic average phonon energy ($\omega_{\text{log}}$) can be computed by using 
\begin{equation}
    \omega_{\text{log}} = \exp\left(\frac{2}{\lambda}\int^{\infty}_{0} d\Omega \text{log}(\Omega) \left(\frac{\alpha^{2}F(\Omega)}{\Omega}\right)\right).
\end{equation}
with
    \begin{equation}
        f = \left(1+\left(\frac{\lambda}{2.46(1+3.8\mu^{*})}\right)\right)^{1/3}\times\left(1+\frac{\lambda^2 (\frac{\omega_2}{\omega_{\text{log}}}-1)}{\lambda^2 +3.31(1+6.3\mu^*)^2}\right).
    \end{equation} 
This $f$ correction factor will be used when the electron-phonon coupling is typically larger than 1.0 where the mean-square frequency ($\omega_2$) is given by
\begin{equation}
    \omega_2 = \sqrt{\frac{2}{\lambda}\int_{0}^{\omega_{\text{max}}}\alpha^2 F(\omega)\omega d\omega}.
\end{equation}
In the main analysis, the superconducting temperature is calculated with the Morel-Anderson pseudopotential set to a practical value of $\mu^* = 0.1$. However, the actual $\mu^*$ values for these materials may vary, leading to different $T_c$ values, as illustrated in the appendix.
\section{Results and Discussions}
\subsection{Crystal structure}
    \begin{figure}[h]
		\centering  
    	\includegraphics[width=9cm]{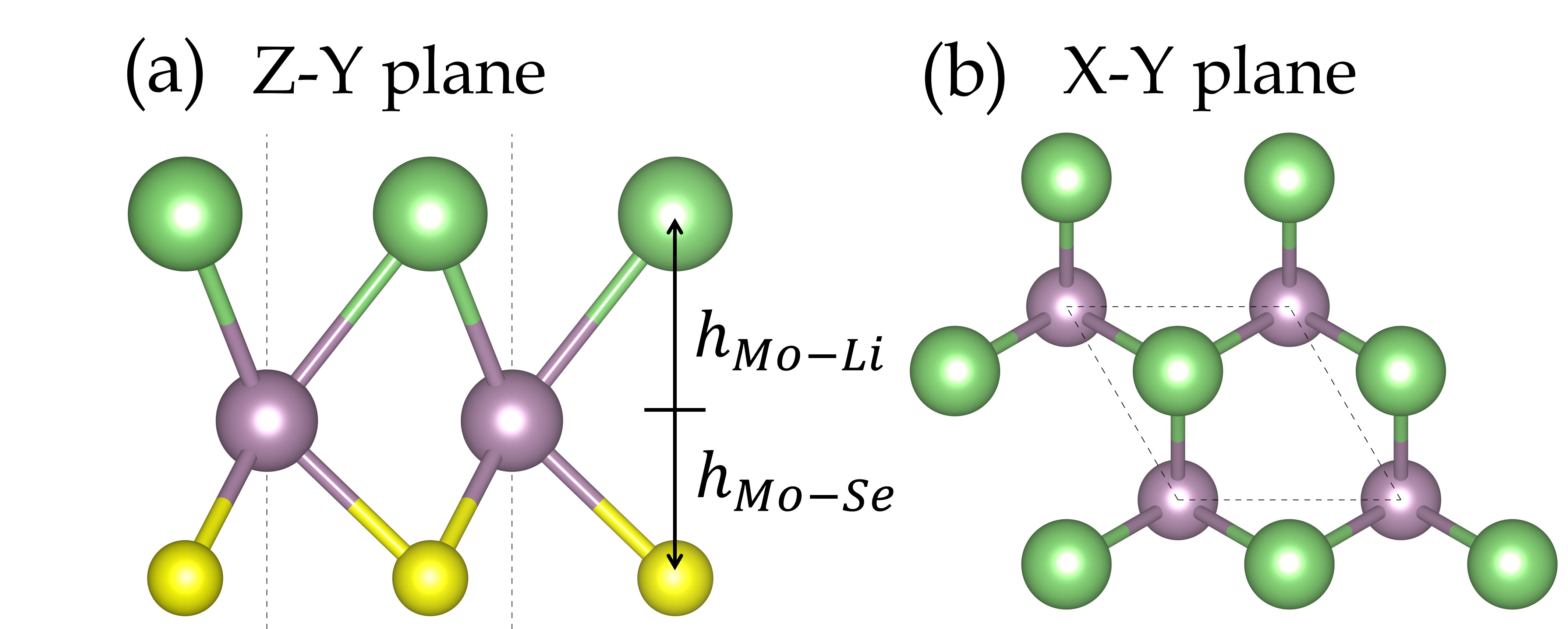}
		\caption{Figures (a) and (b) show the side-view and top-view crystal structures of 2H-MoSeLi, where the purple (Mo), yellow (Se), and green (Li) spheres represent Mo, Se, and Li atoms, respectively.}
		\label{fig:structures}
	\end{figure}
    The hexagonal structure of Janus MoSeLi belongs to the 3D trigonal space group $P3m1$ (No.156), as depicted in Figure~\ref{fig:structures}. In this structure, the transition metal atom, Mo, is located at the Wyckoff position (0,0), while the chalcogenide atom, Se, occupies the Wyckoff positions (1/3, 2/3) in the $x$-$y$ plane, with a single variable lattice constant $a$. The lithium atom can occupy either the (1/3, 2/3) or (2/3, 1/3) positions, depending on whether the material is in the 2H phase or the 1T phase, respectively. Our investigations have shown that the 2H phase is more energetically favorable than the 1T phase. Furthermore, the 1T phase is not dynamically stable, as indicated by the density functional perturbation theory (DFPT) calculations of phonons. Thus, we will focus only the 2H phase of MoSeLi from now on. The magnetic phases, including ferromagnetism and antiferromagnetism (G-type antiferromagnetism (GAF), and C-type antiferromagnetism, (CAF)) of MoSeLi were tested, and we found that the non-magnetic metallic phases are energetically favorable. Along the $z$ axis, the positions of Mo and Se are characterized by two variables: the distance between the Mo and Li sublayers, denoted as $h_{\text{Mo-Li}} = 2.26$ \AA, and the distance between the Mo and Se sublayers, denoted as $h_{\text{Mo-Se}} = 1.72$ \AA.
 
    \begin{figure}[h]
		\centering  
    	\includegraphics[width=6.1cm]{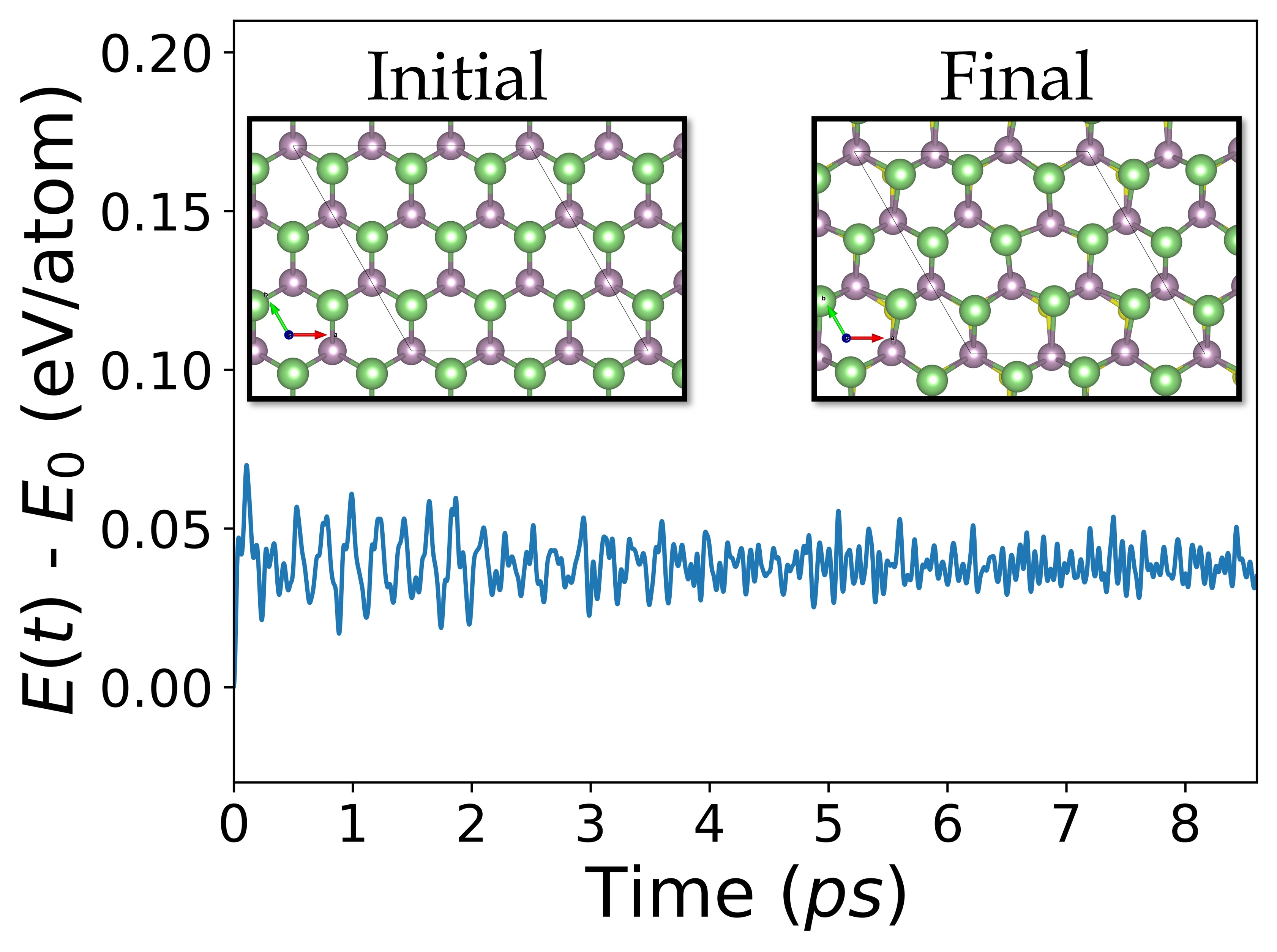}
		\caption{Figures display the NVT ensemble results of the Born-Oppenheimer molecular dynamics (BOMD) simulations for a $3 \times 3 \times 1$ supercell of MoSeLi, consisting of 27 atoms, at 300 K, where we show the initial and final configurations in left and right sub-panels, respectively.}
		\label{fig:MD-analysis}
	\end{figure} 
    By performing molecular dynamics simulations on a $3\times3\times1$ supercell of the hexagonal 2H Janus MoSeLi monolayer (2H-MoSeLi), we have confirmed its thermal stability, which complements the phonon dynamical stability discussed in the following section. The simulations indicate that the atomic configuration remains similar to the initial configuration, with the energy oscillating around the equilibrium energy, as illustrated in Figure~\ref{fig:MD-analysis}.

\subsection{Electronic properties}
    \begin{figure}[h]
		\centering  
    	\includegraphics[width=9.2cm]{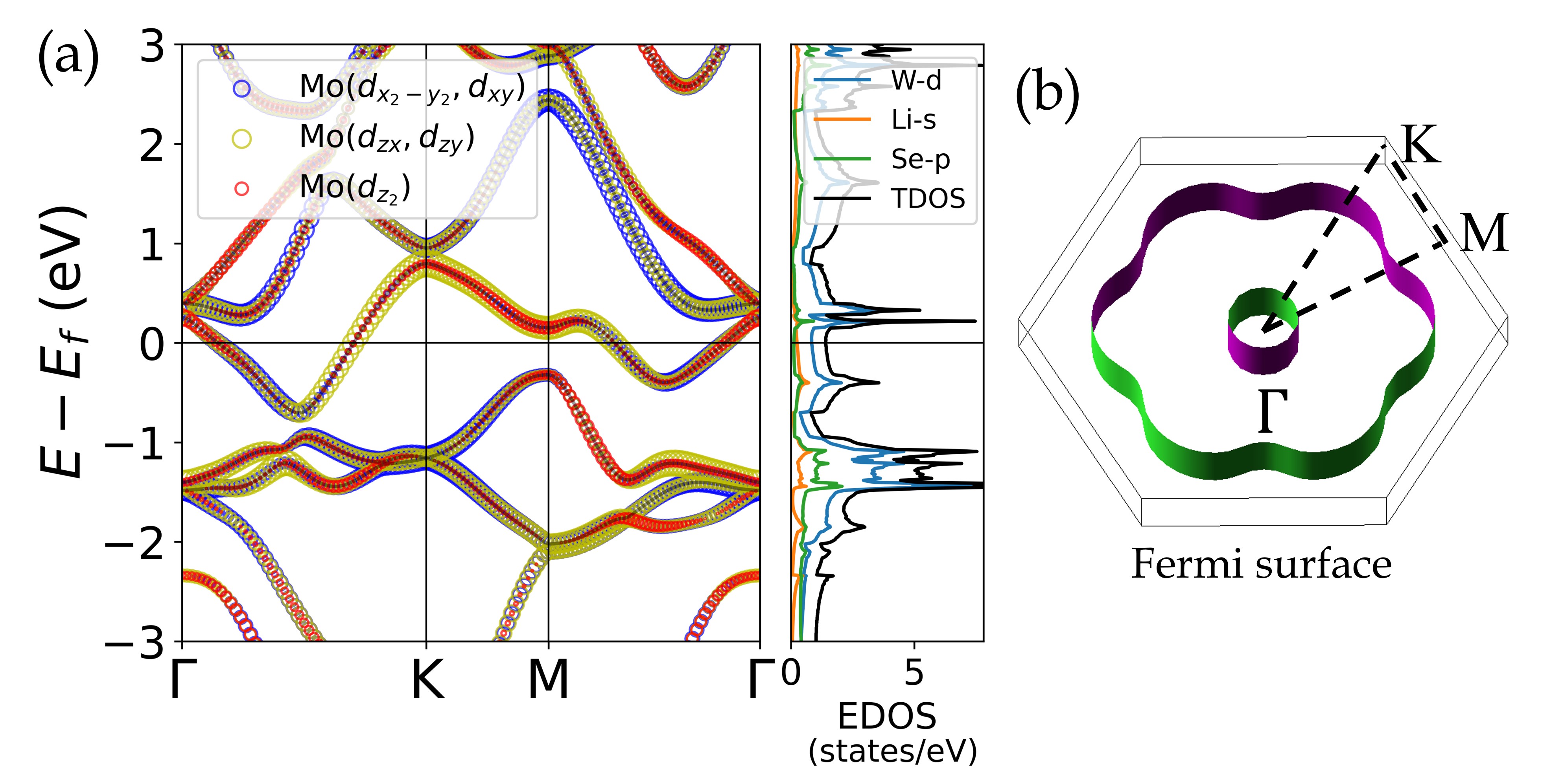}
		\caption{Figures (a) and (b) show the orbital-resolved electronic band structures, the electronic density of states, the orbital-projected density of states, and the Fermi surface of 2H-MoSeLi.}
		\label{fig:electronics}
	\end{figure}
    The electronic structures are presented in Figure~\ref{fig:electronics}. These figures illustrate the characteristic behaviors of electrons near the Fermi level, including the orbital-resolved electronic band structures, the electronic density of states, the orbital-projected density of states, and the Fermi surface in the Brillouin zone. MoSeLi exhibits metallic behavior due to degenerate band crossings at the Fermi level, which mainly arise from contributions of Mo-$d$ orbitals. At the $\Gamma$ point, the Mo-$d$ orbitals can be categorized into $A'(d_{z^2})$, $E'(d_{xy}, d_{x^2-y^2})$, and $E''(d_{yz}, d_{xz})$. From Figure~\ref{fig:electronics} (a), it is evident that MoSeLi exhibits metallic properties due to dominated $d$-band crossings at the Fermi level. The first and second intersections of the bands at the Fermi level occur along the $\Gamma$ to $K$ path, as well as along the $\Gamma$ to $M$ path. These intersections correspond to the inner and outer shells of the Fermi surface, as shown in Figure~\ref{fig:electronics} (b). The orbital-resolved band structure reveals that the outer shell of the Fermi surface is predominantly influenced by the $d_{yz}$ and $d_{xz}$ orbitals, while the inner shell arises from the hybridization of $d_{z^2}$, $d_{yz}$, $d_{xz}$, and $d_{xy}$, $d_{x^2-y^2}$ orbitals.

\subsection{Phonons}
    \begin{figure}[h]
		\centering  
    	\includegraphics[width=8.8cm]{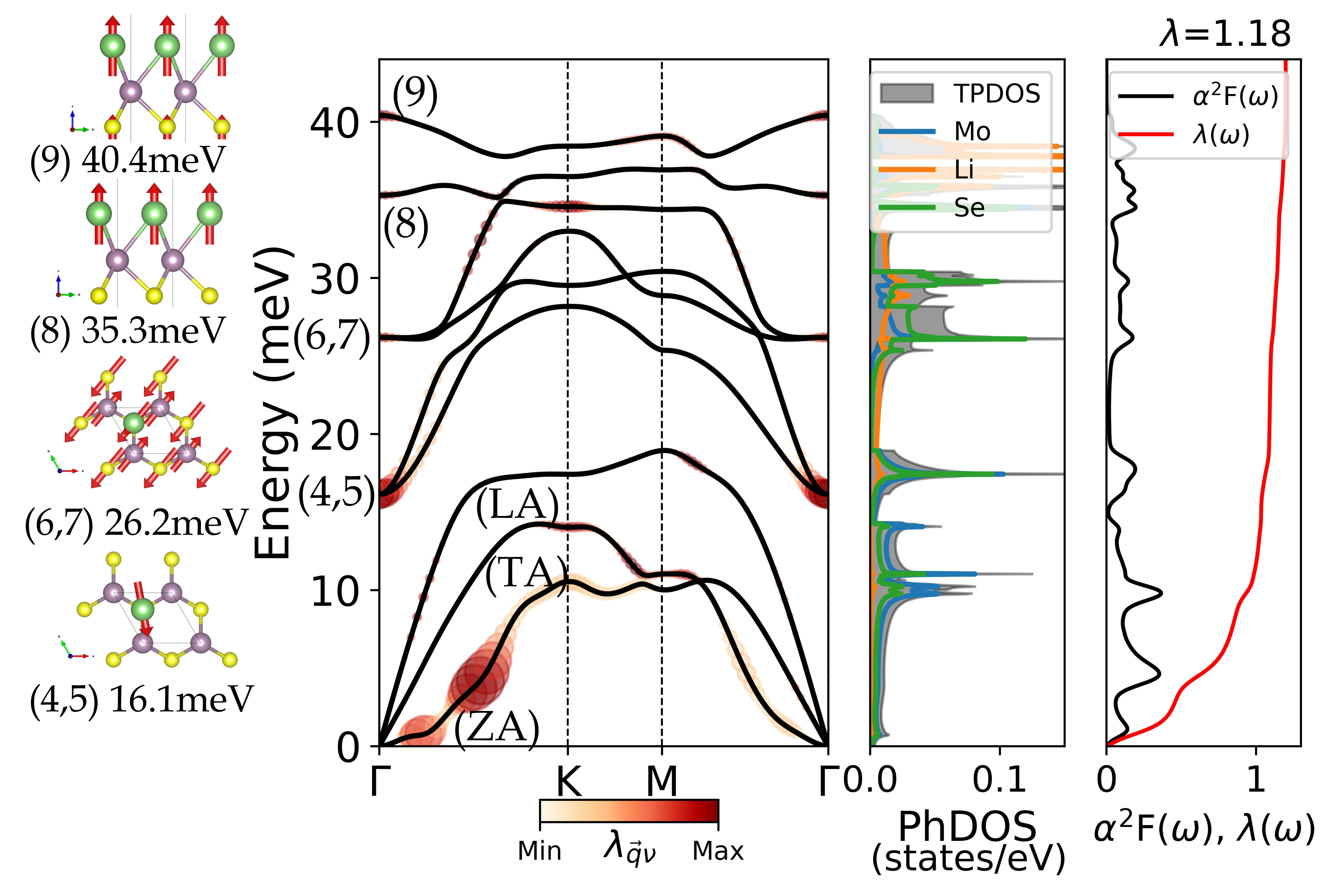}
		\caption{Figures show the weighted electron-phonon coupling (EPC) phonon dispersion, phonon density of states, and the isotropic Eliashberg spectral functions $\alpha^2 F(\omega)$ and $\lambda(\omega)$. Additionally, the figures display the six optical vibrational modes of 2H-MoSeLi at the $\Gamma$ point in the Brillouin zone. The color of the spheres in the figures corresponds to the same information as shown in Figure~\ref{fig:structures}. The frequency eigenvalues of the corresponding optical vibrational modes are provided in Table~\ref{tab:vibrational-modes}.}
		\label{fig:phonons}
    \end{figure}  
    2H-MoSeLi is dynamically stable, as evidenced by the positive-definite phonon spectrum presented in Figure~\ref{fig:phonons}. At the $\Gamma$ point, the phonon vibrational modes, which belong to the $C_{3v} (3m)$ point group, consist of three acoustic and six optical modes. The three acoustic modes are the in-plane longitudinal acoustic (LA) mode, the in-plane transverse acoustic (TA) mode, and the out-of-plane flexural (ZA) mode, as illustrated in the phonon dispersion in Figure~\ref{fig:phonons}. The out-of-plane flexural (ZA) mode exhibits phonon softening between the $\Gamma$ and $K$ points, indicating a small dip in the phonon dispersion. This softening can enhance the electron-phonon coupling at the corresponding wavevector $\boldsymbol{q}$. The optical bands are categorized into two sets of $E$ and $A_{1}$ subgroups. The former includes bands (4,5) associated with in-plane vibrations of Li, and bands (6,7) corresponding to in-plane vibrations of Se and Li, as illustrated in Figure~\ref{fig:phonons}. The latter features bands (8) and (9), where (8) is attributed to out-of-plane vibrations of Li, and (9) is attributed to out-of-plane vibrations of Se and Li, as depicted in Figure~\ref{fig:phonons}. The absence of three-dimensional inversion symmetry results in all modes being both Raman and infrared (IR) active. A summary of these details is provided in Table~\ref{tab:vibrational-modes} and further illustrated in Figure~\ref{fig:phonons}.

\begin{table}[h]
\caption{\label{tab:vibrational-modes}
provides a comprehensive summary of the phonon modes for MoSeLi. It includes details such as band numbers, subgroups, eigenvectors, and whether the modes are infrared (IR) or Raman active, along with their frequencies (meV) at the $\Gamma$ point.}
\begin{ruledtabular}
\begin{tabular}{ccccccc}
\textrm{Band $\nu$}&
\textrm{Subgroup}&
\textrm{Eigenvector}&
\textrm{Active}&
\textrm{MoSeLi}\\
\colrule
		4,5 & $E$ & \text{In-plane Li} & I+R & 16.1  \\
		6,7 & $E$ & \text{In-plane Se, Li} & I+R & 26.2  \\
		8 & $A_{1}$ & \text{Out-plane Li} & I+R & 35.3  \\
		9 & $A_{1}$ & \text{Out-plane Se, Li}  & I+R & 40.4 \\
\end{tabular}
\end{ruledtabular}
\end{table}   

\subsection{Electron-phonon interaction}
    \begin{figure}[h]
		\centering  
    	\includegraphics[width=9cm]{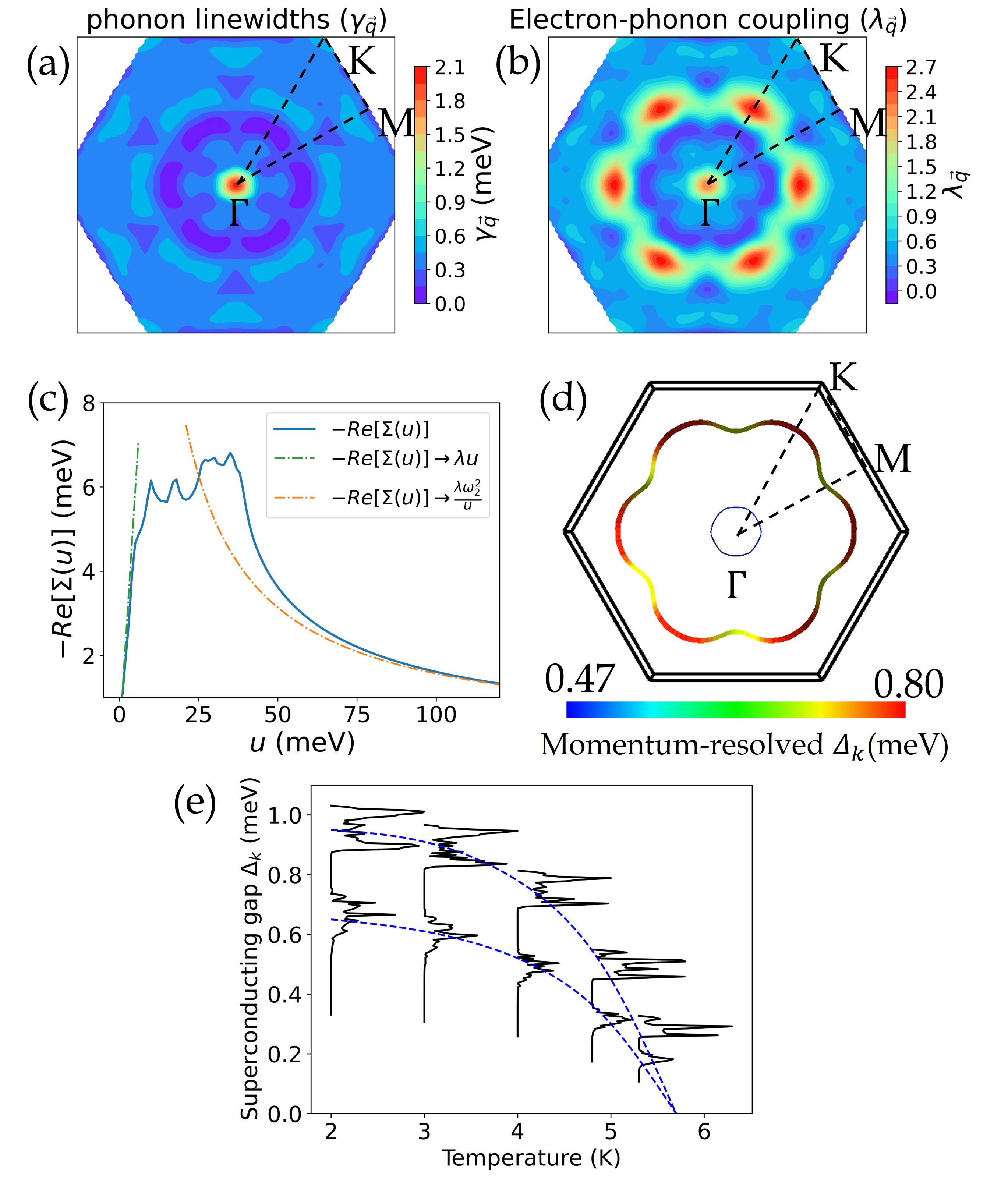}
		\caption{Figures show (a) the total phonon linewidth ($\gamma_{\bold{q}\nu}$), (b) the coutour of electron-phonon coupling ($\lambda_{\bold{q}\nu}$) in the Brillioun zone, (c) the electron self-energy and (d)the two superconducting gaps of MoSeLi.}
		\label{fig:elph}
	\end{figure}
     The electron-phonon interaction is illustrated through the total phonon linewidth ($\gamma_{\bold{q}\nu}$) and the total electron-phonon coupling ($\lambda_{\bold{q}\nu}$) as shown in Figure~\ref{fig:elph} (a) and (b) as functions of the phonon wavevector $\bold{q}$, and summing over bands $\nu$ ranging from 1 to 9. The electron-phonon interaction treats phonons as quasiparticles with a finite lifetime, which is directly related to $\gamma_{\bold{q}\nu}$. Additionally, $\gamma_{\bold{q}\nu}$ also describes the weighted connection between points on the Fermi surfaces mediated by the phonon wavevector $\bold{q}$ of band $\nu$, called Fermi nesting.
     
     From $\gamma_{q \nu}$, it is evident that the strongest connections occur at low $\boldsymbol{q}$ near $\Gamma$, compared to moderate $\boldsymbol{q}$ between $\Gamma$ and $K$, and high $\boldsymbol{q}$ at $M$, as shown in Figure~\ref{fig:elph} (a). At low $\boldsymbol{q}$ near $\Gamma$, the prominent peaks are associated with optical modes ($\nu = 4-9$), as $\gamma_{\bold{q}\nu} \rightarrow 0$ for acoustic modes at long-wavelength limit ($\bold{q} \rightarrow 0$). At moderate $\boldsymbol{q}$ between $\Gamma$ and $K$, the peaks result from both optical and acoustic modes ($\nu = 1-3$), with the ZA mode being particularly significant. At high $\boldsymbol{q}$ at $M$, the peaks are predominantly contributed by the optical mode of $\nu = 9$.

    From $\lambda_{\bold{q}\nu} \propto \frac{\gamma_{\bold{q}\nu}}{\omega^2_{\bold{q}\nu}}$, we can analyze the general behaviors in acoustic and optical modes. Since $\lambda = \sum_{\bold{q}\nu} \lambda_{\bold{q}\nu}$, we would expect high values of $\lambda_{\bold{q}\nu}$ at low $\boldsymbol{q}$ near $\Gamma$ compared to moderate $\boldsymbol{q}$ between $\Gamma$ and $K$, and high $\boldsymbol{q}$ near $M$, given the high values of $\gamma_{q \nu}$. However, specific electronic and phonon structures give rise to noticeable patterns in $\lambda_{\bold{q}\nu}$. At $\Gamma$, $\lambda_{\bold{q}\nu}$ is primarily contributed by optical modes ($\nu = 4-9$), which are typically suppressed by their high frequencies. The acoustic modes ($\nu = 1-3$) contribute less significantly, with $\gamma_{\bold{q}\nu} \rightarrow 0$ and $\omega_{\bold{q}\nu} \rightarrow 0$ as $\bold{q} \rightarrow 0$. Nonetheless, as $\boldsymbol{q}$ increasing near and slightly away from $\Gamma$, the acoustic modes begin to dominate $\lambda_{\bold{q}\nu}$. This is similar to the behavior observed at moderate $\boldsymbol{q}$ between $\Gamma$ and $K$, where the softening of the acoustic ZA mode contributes significantly, as shown in the weighted electron-phonon coupling (EPC) phonon dispersion in Figure~\ref{fig:phonons}. At high $\boldsymbol{q}$ near $M$, there are no special features in the phonons because the high phonon frequencies of mode $\nu=9$ lead to moderate values of $\lambda_{\bold{q}\nu}$. Therefore, $\lambda_{\bold{q}\nu}$ is significantly contributed from the Brillouin zone around the $\Gamma$ point and between $\Gamma$ and $K$, where $\gamma_{\bold{q}\nu}$ is large and the phonon is softening, respectively. The existence of softening phonon plays very important role in strong electron-phonon coupling ($\lambda$) as also observed and mentioned in transition metal chalcogenide hydrides \cite{liu2022two,seeyangnok2024superconductivity}.

    From $\alpha^2 \text{F}(\omega)$, we compute the real part of the electron self-energy due to electron-phonon interactions as illustrated in Figure~\ref{fig:elph} (c). This self energy reflects the cumulative effect integrated across the entire Brillouin zone. Experimental techniques such as angle-resolved photoemission spectroscopy (ARPES) can directly probe the electron self-energy, allowing for direct comparison with our theoretical results, similar to the study of Li-decorated graphene \cite{ludbrook2015evidence}. Figure~\ref{fig:elph} shows that the Eliashberg spectral function features multiple small peaks across various frequency regions, each corresponding to the minor peaks observed in the electron self-energy. Notably, the positions of these peaks in the electron self-energy align closely with the characteristic frequency regions observed in the Eliashberg spectral function. From the electron self-energy depicted in Figure~\ref{fig:elph}, we derive important parameters such as $\lambda$ and $\omega_2$. The parameter $\lambda$ is obtained from the slope of the electron self-energy at low energies, specifically, $-\frac{\partial \text{Re}[\Sigma(u)]}{\partial u} = \lambda$. Additionally, $\omega_2$ is determined from the decay tail of the electron self-energy at high energies, approximated by $- \text{Re}[\Sigma(u)] \approx \frac{\lambda \omega^2_2}{u}$, as shown in Figure~\ref{fig:elph} (c). These limits are indicated with $\lambda = 1.18$ and $\omega_2 = 11.53 \text{ meV}$, as shown by the dashed lines (green and orange) in Figure~\ref{fig:elph} (c).

\subsection{Phonon-mediated superconductivity}
    As shown in Figure~\ref{fig:elph} (e), MoSeLi exhibits two-gap superconductivity. Figure~\ref{fig:elph} (d) depicts the momentum-resolved superconducting gap $\Delta_k$ on the Fermi surface at 4 K. The relatively large $\Delta_k$ observed is associated with the outer Fermi pockets, which are primarily influenced by the Mo $d_{yz,xz}$ states. In contrast, the inner Fermi pockets are a result of the hybridization of the $d_{z^2}$, $d_{yz,xz}$, and $d_{xy,x^2-y^2}$ orbitals, as analyzed in Figure~\ref{fig:electronics}. Figure~\ref{fig:elph} (e) shows the distribution of superconducting gaps ($\Delta_k$) as a function of temperature. At 4 K, $\Delta_k$ ranges from 0.69 to 0.80 meV and from 0.47 to approximately 0.53 meV, displaying anisotropy, which aligns with the momentum-resolved $\Delta_k$ in Figure~\ref{fig:elph} (d). As the temperature increases, $\Delta_k$ values gradually decrease and eventually vanish at 5.7 K. Overall, similar to other materials such as n-doped graphene \cite{margine2014two}, AlB$_2$-based films \cite{zhao2019two}, trilayer LiB$_2$C$_2$ \cite{gao2020strong}, monolayer LiBC \cite{modak2021prediction}, MoSLi \cite{xie2024strong}, and MoSH \cite{liu2022two}, MoSeLi naturally exhibits a two-gap superconducting nature. We report two values of the critical temperature: By using the Allen-Dynes formula ($\lambda = 1.18$ and $\omega_{\text{log}} = 4.0$ meV), T$_c$ (AD) = 4.5 K, and by the closing superconducting gap of Migdal-Eliashberg equations, T$_c$ (ME) = 5.7 K.

    \section{CONCLUSION}
    In summary, we have explored the electronic structure and electron-phonon coupling (EPC) of MoSeLi. Our analysis reveals that the Mo-\emph{d} orbitals are predominant in the electronic density of states (DOS) near the Fermi level. Phonon stability is confirmed by the positive-definite phonon spectrum obtained from DFPT, as shown in both the weighted EPC phonon dispersion and the phonon density of states. Thermal stability is supported by BOMD simulations at 300K. We observe significant softening phonon in low-energy acoustic ZA mode, contributing to a strong EPC (\(\lambda = 1.18\)). Moreover, it is shown that MoSeLi has a unique electronic structure of different orbital contributions between the inner and outer shells of the Fermi surface, which results in two-gap superconductivity. The critical superconducting temperature (\(T_c\)) is estimated to be 4.5 K and 5.7 K from the Allen-Dynes and Migdal-Eliashberg equations, respectively. These results suggest that further theoretical and experimental investigations are warranted to explore superconductivity in lithium-decorated Janus transition-metal chalcogenides.
\begin{acknowledgments}
	This research project is supported by the Second Century Fund (C2F), Chulalongkorn University. We acknowledge the supporting computing infrastructure provided by NSTDA, CU, CUAASC, NSRF via PMUB [B05F650021, B37G660013] (Thailand). URL:www.e-science.in.th. GJA acknowledges funding from the ERC project Hecate. We also acknowledge the supports from the Cirrus UK National Tier-2 HPC Service at EPCC (http://www.cirrus.ac.uk), funded by the University of Edinburgh and EPSRC (EP/P020267/1).
 
\end{acknowledgments}

\section*{Appendix A: Phonon linewidth and electron-phonon coupling.}
\begin{figure}[h]
		\centering  
    	\includegraphics[width=9cm]{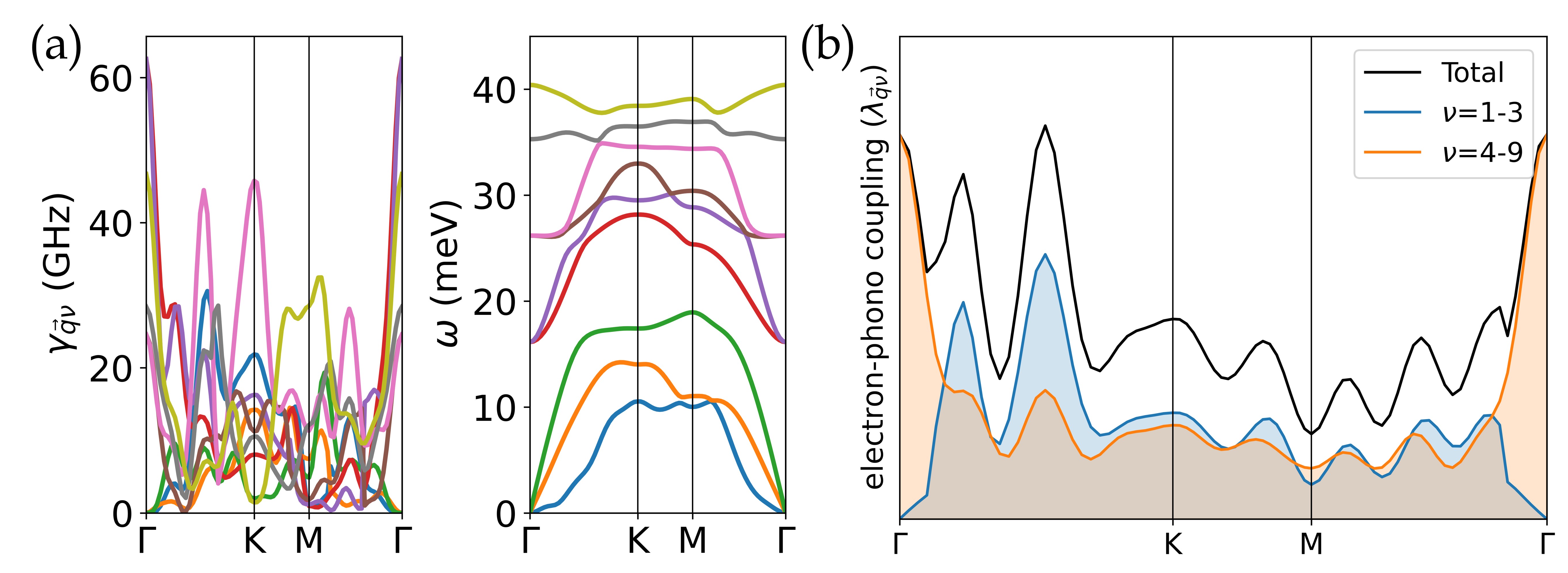}
		\caption{Figures show (a) the phonon linewidths $\gamma_{q\nu}$ and their corresponding phonon modes denoted by different colors, and (b) the electron-phonon coupling $\lambda_{q\nu}$ along symmeytry points within the Brillouin zone.}
		\label{fig:epc}
    \end{figure}

    Both of phonon linewidth ($\gamma_{\boldsymbol{q}\nu}$) and electron-phonon coupling ($\lambda_{\boldsymbol{q}\nu}$) are shown in Figure~\ref{fig:epc}.

\section*{Appendix B: Wanneir interpolation}
\begin{figure}[h]
		\centering  
    	\includegraphics[width=7cm]{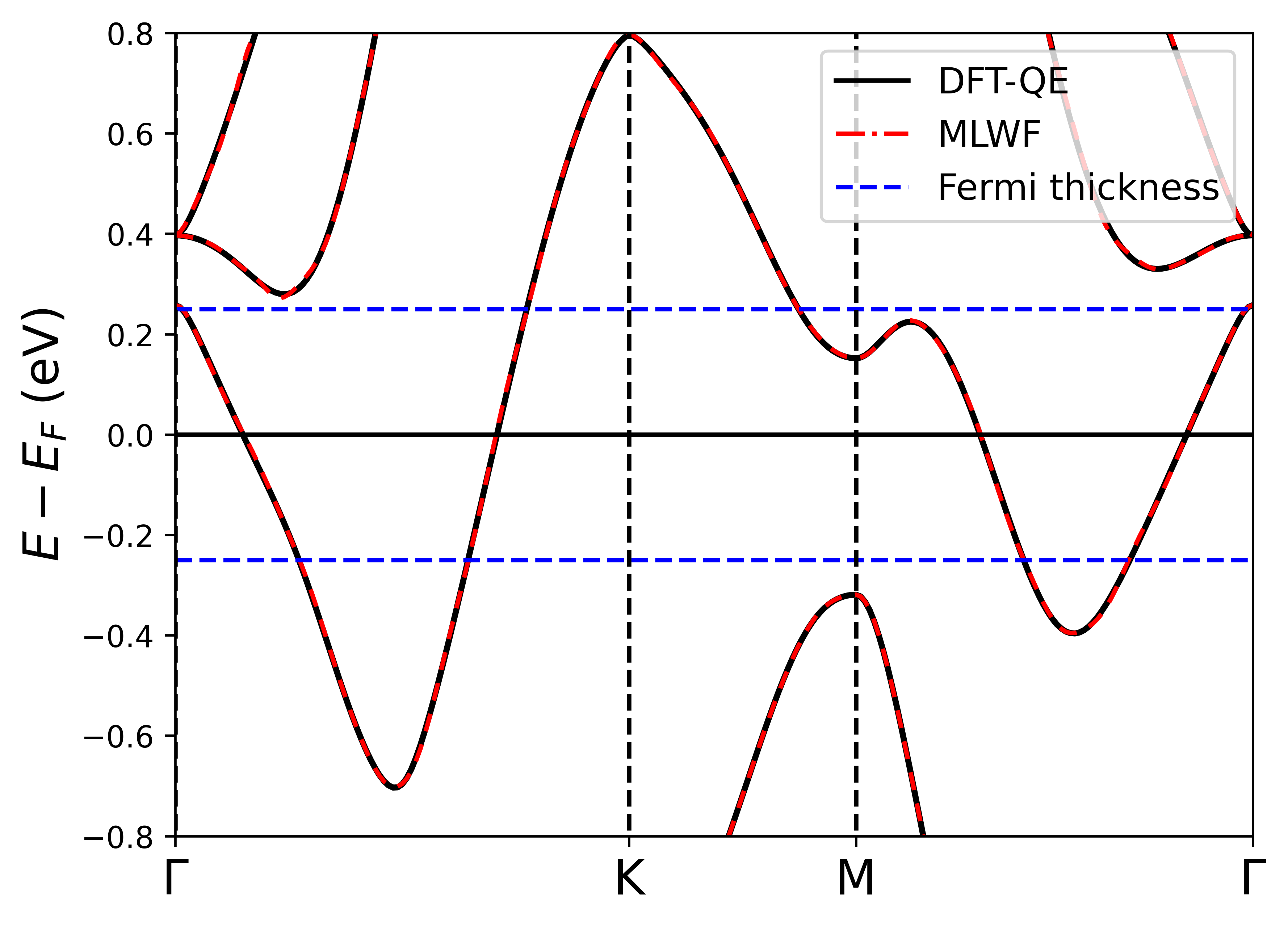}
		\caption{The figures present a comparison between electronic bands computed using standard DFT calculations (black) and those obtained via maximally-localized Wannier functions (dashed red). Additionally, the figures illustrate the Fermi surface around the Fermi level, which is essential for computing the electron-phonon interaction when solving the anisotropic Migdal-Eliashberg equations.}
		\label{fig:epw}
    \end{figure}

The maximally-localized Wannier functions for the investigation of electron-phonon interaction is shown in Figure~\ref{fig:epw}.

as shown in Figure~\ref{fig:epw} within the EPW package.
\section*{Appendix C: The Morel-Anderson pseudopotential ($\mu^*$) dependent T$_c$}
\begin{figure}[h]
		\centering  
    	\includegraphics[width=7cm]{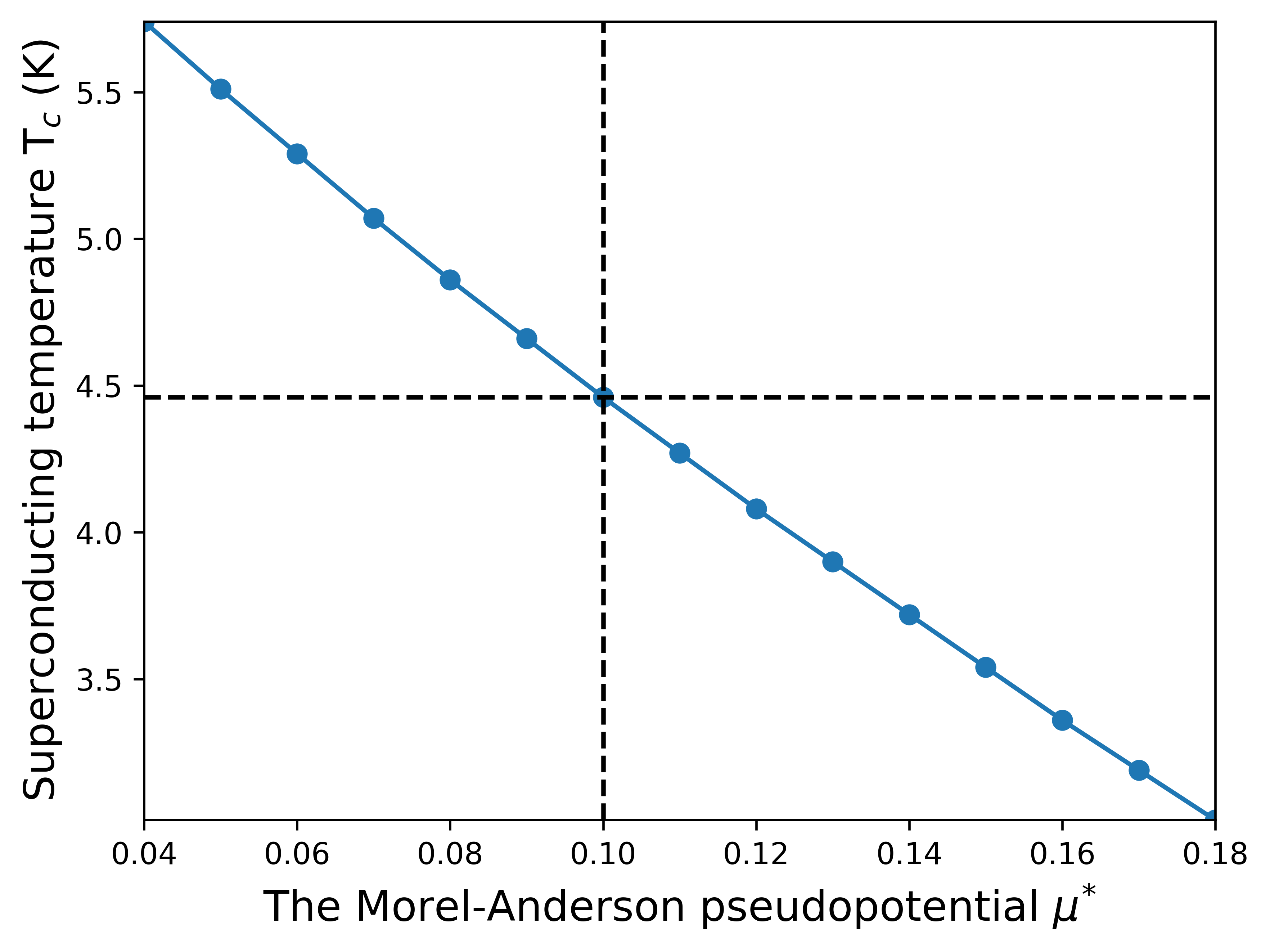}
		\caption{shows the Morel-Anderson pseudopotential ($\mu^*$) dependent T$_c$ of MoSeLi.}
		\label{fig:tc-mu}
    \end{figure}

In the main analysis, the superconducting temperature is calculated with the Morel-Anderson pseudopotential set to a practical value of $\mu^* = 0.1$. However, the actual $\mu^*$ values for these materials may vary, leading to different $T_c$ values, as illustrated in Figure~\ref{fig:tc-mu}.

\bibliography{apssamp}% Produces the bibliography via BibTeX.

%apsrev4-2.bst 2019-01-14 (MD) hand-edited version of apsrev4-1.bst
%Control: key (0)
%Control: author (8) initials jnrlst
%Control: editor formatted (1) identically to author
%Control: production of article title (0) allowed
%Control: page (0) single
%Control: year (1) truncated
%Control: production of eprint (0) enabled
\begin{thebibliography}{54}%
\makeatletter
\providecommand \@ifxundefined [1]{%
 \@ifx{#1\undefined}
}%
\providecommand \@ifnum [1]{%
 \ifnum #1\expandafter \@firstoftwo
 \else \expandafter \@secondoftwo
 \fi
}%
\providecommand \@ifx [1]{%
 \ifx #1\expandafter \@firstoftwo
 \else \expandafter \@secondoftwo
 \fi
}%
\providecommand \natexlab [1]{#1}%
\providecommand \enquote  [1]{``#1''}%
\providecommand \bibnamefont  [1]{#1}%
\providecommand \bibfnamefont [1]{#1}%
\providecommand \citenamefont [1]{#1}%
\providecommand \href@noop [0]{\@secondoftwo}%
\providecommand \href [0]{\begingroup \@sanitize@url \@href}%
\providecommand \@href[1]{\@@startlink{#1}\@@href}%
\providecommand \@@href[1]{\endgroup#1\@@endlink}%
\providecommand \@sanitize@url [0]{\catcode `\\12\catcode `\$12\catcode `\&12\catcode `\#12\catcode `\^12\catcode `\_12\catcode `\%12\relax}%
\providecommand \@@startlink[1]{}%
\providecommand \@@endlink[0]{}%
\providecommand \url  [0]{\begingroup\@sanitize@url \@url }%
\providecommand \@url [1]{\endgroup\@href {#1}{\urlprefix }}%
\providecommand \urlprefix  [0]{URL }%
\providecommand \Eprint [0]{\href }%
\providecommand \doibase [0]{https://doi.org/}%
\providecommand \selectlanguage [0]{\@gobble}%
\providecommand \bibinfo  [0]{\@secondoftwo}%
\providecommand \bibfield  [0]{\@secondoftwo}%
\providecommand \translation [1]{[#1]}%
\providecommand \BibitemOpen [0]{}%
\providecommand \bibitemStop [0]{}%
\providecommand \bibitemNoStop [0]{.\EOS\space}%
\providecommand \EOS [0]{\spacefactor3000\relax}%
\providecommand \BibitemShut  [1]{\csname bibitem#1\endcsname}%
\let\auto@bib@innerbib\@empty
%</preamble>
\bibitem [{\citenamefont {Profeta}\ \emph {et~al.}(2012)\citenamefont {Profeta}, \citenamefont {Calandra},\ and\ \citenamefont {Mauri}}]{profeta2012phonon}%
  \BibitemOpen
  \bibfield  {author} {\bibinfo {author} {\bibfnamefont {G.}~\bibnamefont {Profeta}}, \bibinfo {author} {\bibfnamefont {M.}~\bibnamefont {Calandra}},\ and\ \bibinfo {author} {\bibfnamefont {F.}~\bibnamefont {Mauri}},\ }\bibfield  {title} {\bibinfo {title} {Phonon-mediated superconductivity in graphene by lithium deposition},\ }\href@noop {} {\bibfield  {journal} {\bibinfo  {journal} {Nature physics}\ }\textbf {\bibinfo {volume} {8}},\ \bibinfo {pages} {131} (\bibinfo {year} {2012})}\BibitemShut {NoStop}%
\bibitem [{\citenamefont {Ludbrook}\ \emph {et~al.}(2015)\citenamefont {Ludbrook}, \citenamefont {Levy}, \citenamefont {Nigge}, \citenamefont {Zonno}, \citenamefont {Schneider}, \citenamefont {Dvorak}, \citenamefont {Veenstra}, \citenamefont {Zhdanovich}, \citenamefont {Wong}, \citenamefont {Dosanjh} \emph {et~al.}}]{ludbrook2015evidence}%
  \BibitemOpen
  \bibfield  {author} {\bibinfo {author} {\bibfnamefont {B.}~\bibnamefont {Ludbrook}}, \bibinfo {author} {\bibfnamefont {G.}~\bibnamefont {Levy}}, \bibinfo {author} {\bibfnamefont {P.}~\bibnamefont {Nigge}}, \bibinfo {author} {\bibfnamefont {M.}~\bibnamefont {Zonno}}, \bibinfo {author} {\bibfnamefont {M.}~\bibnamefont {Schneider}}, \bibinfo {author} {\bibfnamefont {D.}~\bibnamefont {Dvorak}}, \bibinfo {author} {\bibfnamefont {C.}~\bibnamefont {Veenstra}}, \bibinfo {author} {\bibfnamefont {S.}~\bibnamefont {Zhdanovich}}, \bibinfo {author} {\bibfnamefont {D.}~\bibnamefont {Wong}}, \bibinfo {author} {\bibfnamefont {P.}~\bibnamefont {Dosanjh}}, \emph {et~al.},\ }\bibfield  {title} {\bibinfo {title} {Evidence for superconductivity in li-decorated monolayer graphene},\ }\href@noop {} {\bibfield  {journal} {\bibinfo  {journal} {Proceedings of the National Academy of Sciences}\ }\textbf {\bibinfo {volume} {112}},\ \bibinfo {pages} {11795} (\bibinfo {year} {2015})}\BibitemShut {NoStop}%
\bibitem [{\citenamefont {Benaddi}\ \emph {et~al.}(2024)\citenamefont {Benaddi}, \citenamefont {Elomrani}, \citenamefont {Lamhani}, \citenamefont {Oukahou}, \citenamefont {Maymoun}, \citenamefont {Fatihi},\ and\ \citenamefont {Hasnaoui}}]{benaddi2024lithium}%
  \BibitemOpen
  \bibfield  {author} {\bibinfo {author} {\bibfnamefont {A.}~\bibnamefont {Benaddi}}, \bibinfo {author} {\bibfnamefont {A.}~\bibnamefont {Elomrani}}, \bibinfo {author} {\bibfnamefont {M.}~\bibnamefont {Lamhani}}, \bibinfo {author} {\bibfnamefont {S.}~\bibnamefont {Oukahou}}, \bibinfo {author} {\bibfnamefont {M.}~\bibnamefont {Maymoun}}, \bibinfo {author} {\bibfnamefont {M.~Y.}\ \bibnamefont {Fatihi}},\ and\ \bibinfo {author} {\bibfnamefont {A.}~\bibnamefont {Hasnaoui}},\ }\bibfield  {title} {\bibinfo {title} {Lithium decorated 2d orthorhombic (o)-b 2 x 2 monolayers for hydrogen storage: first principles calculations},\ }\href@noop {} {\bibfield  {journal} {\bibinfo  {journal} {Sustainable Energy \& Fuels}\ }\textbf {\bibinfo {volume} {8}},\ \bibinfo {pages} {1719} (\bibinfo {year} {2024})}\BibitemShut {NoStop}%
\bibitem [{\citenamefont {Jiang}\ \emph {et~al.}(2023)\citenamefont {Jiang}, \citenamefont {Xu}, \citenamefont {Munroe},\ and\ \citenamefont {Xie}}]{jiang2023lithium}%
  \BibitemOpen
  \bibfield  {author} {\bibinfo {author} {\bibfnamefont {M.}~\bibnamefont {Jiang}}, \bibinfo {author} {\bibfnamefont {J.}~\bibnamefont {Xu}}, \bibinfo {author} {\bibfnamefont {P.}~\bibnamefont {Munroe}},\ and\ \bibinfo {author} {\bibfnamefont {Z.-H.}\ \bibnamefont {Xie}},\ }\bibfield  {title} {\bibinfo {title} {Lithium-decorated sib monolayer for reversible hydrogen storage: High-capacity realization through strain engineering},\ }\href@noop {} {\bibfield  {journal} {\bibinfo  {journal} {Applied Surface Science}\ }\textbf {\bibinfo {volume} {618}},\ \bibinfo {pages} {156707} (\bibinfo {year} {2023})}\BibitemShut {NoStop}%
\bibitem [{\citenamefont {Zhang}\ and\ \citenamefont {Guo}(2024)}]{zhang2024li}%
  \BibitemOpen
  \bibfield  {author} {\bibinfo {author} {\bibfnamefont {Y.-F.}\ \bibnamefont {Zhang}}\ and\ \bibinfo {author} {\bibfnamefont {J.}~\bibnamefont {Guo}},\ }\bibfield  {title} {\bibinfo {title} {Li-decorated 2d irida-graphene as a potential hydrogen storage material: A dispersion-corrected density functional theory calculations},\ }\href@noop {} {\bibfield  {journal} {\bibinfo  {journal} {International Journal of Hydrogen Energy}\ }\textbf {\bibinfo {volume} {50}},\ \bibinfo {pages} {1004} (\bibinfo {year} {2024})}\BibitemShut {NoStop}%
\bibitem [{\citenamefont {Haldar}\ \emph {et~al.}(2017)\citenamefont {Haldar}, \citenamefont {Mukherjee}, \citenamefont {Ahmed},\ and\ \citenamefont {Singh}}]{haldar2017first}%
  \BibitemOpen
  \bibfield  {author} {\bibinfo {author} {\bibfnamefont {S.}~\bibnamefont {Haldar}}, \bibinfo {author} {\bibfnamefont {S.}~\bibnamefont {Mukherjee}}, \bibinfo {author} {\bibfnamefont {F.}~\bibnamefont {Ahmed}},\ and\ \bibinfo {author} {\bibfnamefont {C.~V.}\ \bibnamefont {Singh}},\ }\bibfield  {title} {\bibinfo {title} {A first principles study of hydrogen storage in lithium decorated defective phosphorene},\ }\href@noop {} {\bibfield  {journal} {\bibinfo  {journal} {international journal of hydrogen energy}\ }\textbf {\bibinfo {volume} {42}},\ \bibinfo {pages} {23018} (\bibinfo {year} {2017})}\BibitemShut {NoStop}%
\bibitem [{\citenamefont {Dewangan}\ \emph {et~al.}(2023)\citenamefont {Dewangan}, \citenamefont {Mahamiya}, \citenamefont {Shukla},\ and\ \citenamefont {Chakraborty}}]{dewangan2023lithium}%
  \BibitemOpen
  \bibfield  {author} {\bibinfo {author} {\bibfnamefont {J.}~\bibnamefont {Dewangan}}, \bibinfo {author} {\bibfnamefont {V.}~\bibnamefont {Mahamiya}}, \bibinfo {author} {\bibfnamefont {A.}~\bibnamefont {Shukla}},\ and\ \bibinfo {author} {\bibfnamefont {B.}~\bibnamefont {Chakraborty}},\ }\bibfield  {title} {\bibinfo {title} {Lithium decorated psi-graphene as a potential hydrogen storage material: density functional theory investigations},\ }\href@noop {} {\bibfield  {journal} {\bibinfo  {journal} {International Journal of Hydrogen Energy}\ }\textbf {\bibinfo {volume} {48}},\ \bibinfo {pages} {37908} (\bibinfo {year} {2023})}\BibitemShut {NoStop}%
\bibitem [{\citenamefont {Tang}\ and\ \citenamefont {Kou}(2022)}]{tang20222d}%
  \BibitemOpen
  \bibfield  {author} {\bibinfo {author} {\bibfnamefont {X.}~\bibnamefont {Tang}}\ and\ \bibinfo {author} {\bibfnamefont {L.}~\bibnamefont {Kou}},\ }\bibfield  {title} {\bibinfo {title} {2d janus transition metal dichalcogenides: Properties and applications},\ }\href@noop {} {\bibfield  {journal} {\bibinfo  {journal} {physica status solidi (b)}\ }\textbf {\bibinfo {volume} {259}},\ \bibinfo {pages} {2100562} (\bibinfo {year} {2022})}\BibitemShut {NoStop}%
\bibitem [{\citenamefont {Zhang}\ \emph {et~al.}(2022)\citenamefont {Zhang}, \citenamefont {Xia}, \citenamefont {Li}, \citenamefont {Li}, \citenamefont {Fu}, \citenamefont {Cheng},\ and\ \citenamefont {Pan}}]{zhang2022janus}%
  \BibitemOpen
  \bibfield  {author} {\bibinfo {author} {\bibfnamefont {L.}~\bibnamefont {Zhang}}, \bibinfo {author} {\bibfnamefont {Y.}~\bibnamefont {Xia}}, \bibinfo {author} {\bibfnamefont {X.}~\bibnamefont {Li}}, \bibinfo {author} {\bibfnamefont {L.}~\bibnamefont {Li}}, \bibinfo {author} {\bibfnamefont {X.}~\bibnamefont {Fu}}, \bibinfo {author} {\bibfnamefont {J.}~\bibnamefont {Cheng}},\ and\ \bibinfo {author} {\bibfnamefont {R.}~\bibnamefont {Pan}},\ }\bibfield  {title} {\bibinfo {title} {Janus two-dimensional transition metal dichalcogenides},\ }\href@noop {} {\bibfield  {journal} {\bibinfo  {journal} {Journal of Applied Physics}\ }\textbf {\bibinfo {volume} {131}} (\bibinfo {year} {2022})}\BibitemShut {NoStop}%
\bibitem [{\citenamefont {Angeli}\ \emph {et~al.}(2022)\citenamefont {Angeli}, \citenamefont {Schleder},\ and\ \citenamefont {Kaxiras}}]{angeli2022twistronics}%
  \BibitemOpen
  \bibfield  {author} {\bibinfo {author} {\bibfnamefont {M.}~\bibnamefont {Angeli}}, \bibinfo {author} {\bibfnamefont {G.~R.}\ \bibnamefont {Schleder}},\ and\ \bibinfo {author} {\bibfnamefont {E.}~\bibnamefont {Kaxiras}},\ }\bibfield  {title} {\bibinfo {title} {Twistronics of janus transition metal dichalcogenide bilayers},\ }\href@noop {} {\bibfield  {journal} {\bibinfo  {journal} {Physical Review B}\ }\textbf {\bibinfo {volume} {106}},\ \bibinfo {pages} {235159} (\bibinfo {year} {2022})}\BibitemShut {NoStop}%
\bibitem [{\citenamefont {He}\ and\ \citenamefont {Li}(2018)}]{he2018two}%
  \BibitemOpen
  \bibfield  {author} {\bibinfo {author} {\bibfnamefont {J.}~\bibnamefont {He}}\ and\ \bibinfo {author} {\bibfnamefont {S.}~\bibnamefont {Li}},\ }\bibfield  {title} {\bibinfo {title} {Two-dimensional janus transition-metal dichalcogenides with intrinsic ferromagnetism and half-metallicity},\ }\href@noop {} {\bibfield  {journal} {\bibinfo  {journal} {Computational Materials Science}\ }\textbf {\bibinfo {volume} {152}},\ \bibinfo {pages} {151} (\bibinfo {year} {2018})}\BibitemShut {NoStop}%
\bibitem [{\citenamefont {Yeh}(2020)}]{yeh2020computational}%
  \BibitemOpen
  \bibfield  {author} {\bibinfo {author} {\bibfnamefont {C.-H.}\ \bibnamefont {Yeh}},\ }\bibfield  {title} {\bibinfo {title} {Computational study of janus transition metal dichalcogenide monolayers for acetone gas sensing},\ }\href@noop {} {\bibfield  {journal} {\bibinfo  {journal} {ACS omega}\ }\textbf {\bibinfo {volume} {5}},\ \bibinfo {pages} {31398} (\bibinfo {year} {2020})}\BibitemShut {NoStop}%
\bibitem [{\citenamefont {Yin}\ \emph {et~al.}(2021)\citenamefont {Yin}, \citenamefont {Tan}, \citenamefont {Ding}, \citenamefont {Wen}, \citenamefont {Li}, \citenamefont {Teobaldi},\ and\ \citenamefont {Liu}}]{yin2021recent}%
  \BibitemOpen
  \bibfield  {author} {\bibinfo {author} {\bibfnamefont {W.-J.}\ \bibnamefont {Yin}}, \bibinfo {author} {\bibfnamefont {H.-J.}\ \bibnamefont {Tan}}, \bibinfo {author} {\bibfnamefont {P.-J.}\ \bibnamefont {Ding}}, \bibinfo {author} {\bibfnamefont {B.}~\bibnamefont {Wen}}, \bibinfo {author} {\bibfnamefont {X.-B.}\ \bibnamefont {Li}}, \bibinfo {author} {\bibfnamefont {G.}~\bibnamefont {Teobaldi}},\ and\ \bibinfo {author} {\bibfnamefont {L.-M.}\ \bibnamefont {Liu}},\ }\bibfield  {title} {\bibinfo {title} {Recent advances in low-dimensional janus materials: theoretical and simulation perspectives},\ }\href@noop {} {\bibfield  {journal} {\bibinfo  {journal} {Materials Advances}\ }\textbf {\bibinfo {volume} {2}},\ \bibinfo {pages} {7543} (\bibinfo {year} {2021})}\BibitemShut {NoStop}%
\bibitem [{\citenamefont {Zhang}\ \emph {et~al.}(2013)\citenamefont {Zhang}, \citenamefont {Yu}, \citenamefont {Yang}, \citenamefont {Xie}, \citenamefont {Peng},\ and\ \citenamefont {Liu}}]{zhang2013janus}%
  \BibitemOpen
  \bibfield  {author} {\bibinfo {author} {\bibfnamefont {L.}~\bibnamefont {Zhang}}, \bibinfo {author} {\bibfnamefont {J.}~\bibnamefont {Yu}}, \bibinfo {author} {\bibfnamefont {M.}~\bibnamefont {Yang}}, \bibinfo {author} {\bibfnamefont {Q.}~\bibnamefont {Xie}}, \bibinfo {author} {\bibfnamefont {H.}~\bibnamefont {Peng}},\ and\ \bibinfo {author} {\bibfnamefont {Z.}~\bibnamefont {Liu}},\ }\bibfield  {title} {\bibinfo {title} {Janus graphene from asymmetric two-dimensional chemistry},\ }\href@noop {} {\bibfield  {journal} {\bibinfo  {journal} {Nature Communications}\ }\textbf {\bibinfo {volume} {4}},\ \bibinfo {pages} {1443} (\bibinfo {year} {2013})}\BibitemShut {NoStop}%
\bibitem [{\citenamefont {Trivedi}\ \emph {et~al.}(2020)\citenamefont {Trivedi}, \citenamefont {Turgut}, \citenamefont {Qin}, \citenamefont {Sayyad}, \citenamefont {Hajra}, \citenamefont {Howell}, \citenamefont {Liu}, \citenamefont {Yang}, \citenamefont {Patoary}, \citenamefont {Li} \emph {et~al.}}]{trivedi2020room}%
  \BibitemOpen
  \bibfield  {author} {\bibinfo {author} {\bibfnamefont {D.~B.}\ \bibnamefont {Trivedi}}, \bibinfo {author} {\bibfnamefont {G.}~\bibnamefont {Turgut}}, \bibinfo {author} {\bibfnamefont {Y.}~\bibnamefont {Qin}}, \bibinfo {author} {\bibfnamefont {M.~Y.}\ \bibnamefont {Sayyad}}, \bibinfo {author} {\bibfnamefont {D.}~\bibnamefont {Hajra}}, \bibinfo {author} {\bibfnamefont {M.}~\bibnamefont {Howell}}, \bibinfo {author} {\bibfnamefont {L.}~\bibnamefont {Liu}}, \bibinfo {author} {\bibfnamefont {S.}~\bibnamefont {Yang}}, \bibinfo {author} {\bibfnamefont {N.~H.}\ \bibnamefont {Patoary}}, \bibinfo {author} {\bibfnamefont {H.}~\bibnamefont {Li}}, \emph {et~al.},\ }\bibfield  {title} {\bibinfo {title} {Room-temperature synthesis of 2d janus crystals and their heterostructures},\ }\href@noop {} {\bibfield  {journal} {\bibinfo  {journal} {Advanced materials}\ }\textbf {\bibinfo {volume} {32}},\ \bibinfo {pages} {2006320} (\bibinfo {year} {2020})}\BibitemShut {NoStop}%
\bibitem [{\citenamefont {Lu}\ \emph {et~al.}(2017)\citenamefont {Lu}, \citenamefont {Zhu}, \citenamefont {Xiao}, \citenamefont {Chuu}, \citenamefont {Han}, \citenamefont {Chiu}, \citenamefont {Cheng}, \citenamefont {Yang}, \citenamefont {Wei}, \citenamefont {Yang} \emph {et~al.}}]{lu2017janus}%
  \BibitemOpen
  \bibfield  {author} {\bibinfo {author} {\bibfnamefont {A.-Y.}\ \bibnamefont {Lu}}, \bibinfo {author} {\bibfnamefont {H.}~\bibnamefont {Zhu}}, \bibinfo {author} {\bibfnamefont {J.}~\bibnamefont {Xiao}}, \bibinfo {author} {\bibfnamefont {C.-P.}\ \bibnamefont {Chuu}}, \bibinfo {author} {\bibfnamefont {Y.}~\bibnamefont {Han}}, \bibinfo {author} {\bibfnamefont {M.-H.}\ \bibnamefont {Chiu}}, \bibinfo {author} {\bibfnamefont {C.-C.}\ \bibnamefont {Cheng}}, \bibinfo {author} {\bibfnamefont {C.-W.}\ \bibnamefont {Yang}}, \bibinfo {author} {\bibfnamefont {K.-H.}\ \bibnamefont {Wei}}, \bibinfo {author} {\bibfnamefont {Y.}~\bibnamefont {Yang}}, \emph {et~al.},\ }\bibfield  {title} {\bibinfo {title} {Janus monolayers of transition metal dichalcogenides},\ }\href@noop {} {\bibfield  {journal} {\bibinfo  {journal} {Nature nanotechnology}\ }\textbf {\bibinfo {volume} {12}},\ \bibinfo {pages} {744} (\bibinfo {year} {2017})}\BibitemShut {NoStop}%
\bibitem [{\citenamefont {Sant}\ \emph {et~al.}(2020)\citenamefont {Sant}, \citenamefont {Gay}, \citenamefont {Marty}, \citenamefont {Lisi}, \citenamefont {Harrabi}, \citenamefont {Vergnaud}, \citenamefont {Dau}, \citenamefont {Weng}, \citenamefont {Coraux}, \citenamefont {Gauthier} \emph {et~al.}}]{sant2020synthesis}%
  \BibitemOpen
  \bibfield  {author} {\bibinfo {author} {\bibfnamefont {R.}~\bibnamefont {Sant}}, \bibinfo {author} {\bibfnamefont {M.}~\bibnamefont {Gay}}, \bibinfo {author} {\bibfnamefont {A.}~\bibnamefont {Marty}}, \bibinfo {author} {\bibfnamefont {S.}~\bibnamefont {Lisi}}, \bibinfo {author} {\bibfnamefont {R.}~\bibnamefont {Harrabi}}, \bibinfo {author} {\bibfnamefont {C.}~\bibnamefont {Vergnaud}}, \bibinfo {author} {\bibfnamefont {M.~T.}\ \bibnamefont {Dau}}, \bibinfo {author} {\bibfnamefont {X.}~\bibnamefont {Weng}}, \bibinfo {author} {\bibfnamefont {J.}~\bibnamefont {Coraux}}, \bibinfo {author} {\bibfnamefont {N.}~\bibnamefont {Gauthier}}, \emph {et~al.},\ }\bibfield  {title} {\bibinfo {title} {Synthesis of epitaxial monolayer janus sptse},\ }\href@noop {} {\bibfield  {journal} {\bibinfo  {journal} {npj 2D Materials and Applications}\ }\textbf {\bibinfo {volume} {4}},\ \bibinfo {pages} {41} (\bibinfo {year} {2020})}\BibitemShut {NoStop}%
\bibitem [{\citenamefont {Liu}\ \emph {et~al.}(2022)\citenamefont {Liu}, \citenamefont {Zheng}, \citenamefont {Li}, \citenamefont {Si}, \citenamefont {Wei}, \citenamefont {Zhang},\ and\ \citenamefont {Wang}}]{liu2022two}%
  \BibitemOpen
  \bibfield  {author} {\bibinfo {author} {\bibfnamefont {P.-F.}\ \bibnamefont {Liu}}, \bibinfo {author} {\bibfnamefont {F.}~\bibnamefont {Zheng}}, \bibinfo {author} {\bibfnamefont {J.}~\bibnamefont {Li}}, \bibinfo {author} {\bibfnamefont {J.-G.}\ \bibnamefont {Si}}, \bibinfo {author} {\bibfnamefont {L.}~\bibnamefont {Wei}}, \bibinfo {author} {\bibfnamefont {J.}~\bibnamefont {Zhang}},\ and\ \bibinfo {author} {\bibfnamefont {B.-T.}\ \bibnamefont {Wang}},\ }\bibfield  {title} {\bibinfo {title} {Two-gap superconductivity in a janus mosh monolayer},\ }\href@noop {} {\bibfield  {journal} {\bibinfo  {journal} {Physical Review B}\ }\textbf {\bibinfo {volume} {105}},\ \bibinfo {pages} {245420} (\bibinfo {year} {2022})}\BibitemShut {NoStop}%
\bibitem [{\citenamefont {Ku}\ \emph {et~al.}(2023)\citenamefont {Ku}, \citenamefont {Yan}, \citenamefont {Si}, \citenamefont {Zhu}, \citenamefont {Wang}, \citenamefont {Wei}, \citenamefont {Pang}, \citenamefont {Li},\ and\ \citenamefont {Zhou}}]{ku2023ab}%
  \BibitemOpen
  \bibfield  {author} {\bibinfo {author} {\bibfnamefont {R.}~\bibnamefont {Ku}}, \bibinfo {author} {\bibfnamefont {L.}~\bibnamefont {Yan}}, \bibinfo {author} {\bibfnamefont {J.-G.}\ \bibnamefont {Si}}, \bibinfo {author} {\bibfnamefont {S.}~\bibnamefont {Zhu}}, \bibinfo {author} {\bibfnamefont {B.-T.}\ \bibnamefont {Wang}}, \bibinfo {author} {\bibfnamefont {Y.}~\bibnamefont {Wei}}, \bibinfo {author} {\bibfnamefont {K.}~\bibnamefont {Pang}}, \bibinfo {author} {\bibfnamefont {W.}~\bibnamefont {Li}},\ and\ \bibinfo {author} {\bibfnamefont {L.}~\bibnamefont {Zhou}},\ }\bibfield  {title} {\bibinfo {title} {Ab initio investigation of charge density wave and superconductivity in two-dimensional janus 2 h/1 t-mosh monolayers},\ }\href@noop {} {\bibfield  {journal} {\bibinfo  {journal} {Physical Review B}\ }\textbf {\bibinfo {volume} {107}},\ \bibinfo {pages} {064508} (\bibinfo {year} {2023})}\BibitemShut {NoStop}%
\bibitem [{\citenamefont {Seeyangnok}\ \emph {et~al.}(2024{\natexlab{a}})\citenamefont {Seeyangnok}, \citenamefont {Ul~Hassan}, \citenamefont {Pinsook},\ and\ \citenamefont {Ackland}}]{seeyangnok2024superconductivity}%
  \BibitemOpen
  \bibfield  {author} {\bibinfo {author} {\bibfnamefont {J.}~\bibnamefont {Seeyangnok}}, \bibinfo {author} {\bibfnamefont {M.~M.}\ \bibnamefont {Ul~Hassan}}, \bibinfo {author} {\bibfnamefont {U.}~\bibnamefont {Pinsook}},\ and\ \bibinfo {author} {\bibfnamefont {G.}~\bibnamefont {Ackland}},\ }\bibfield  {title} {\bibinfo {title} {Superconductivity and electron self-energy in tungsten-sulfur-hydride monolayer},\ }\href@noop {} {\bibfield  {journal} {\bibinfo  {journal} {2D Materials}\ }\textbf {\bibinfo {volume} {11}},\ \bibinfo {pages} {025020} (\bibinfo {year} {2024}{\natexlab{a}})}\BibitemShut {NoStop}%
\bibitem [{\citenamefont {Seeyangnok}\ \emph {et~al.}(2024{\natexlab{b}})\citenamefont {Seeyangnok}, \citenamefont {Pinsook},\ and\ \citenamefont {Ackland}}]{seeyangnok2024superconductivitywseh}%
  \BibitemOpen
  \bibfield  {author} {\bibinfo {author} {\bibfnamefont {J.}~\bibnamefont {Seeyangnok}}, \bibinfo {author} {\bibfnamefont {U.}~\bibnamefont {Pinsook}},\ and\ \bibinfo {author} {\bibfnamefont {G.~J.}\ \bibnamefont {Ackland}},\ }\bibfield  {title} {\bibinfo {title} {Superconductivity and strain-enhanced phase stability of janus tungsten chalcogenide hydride monolayers},\ }\href@noop {} {\bibfield  {journal} {\bibinfo  {journal} {Physical Review B}\ }\textbf {\bibinfo {volume} {110}},\ \bibinfo {pages} {195408} (\bibinfo {year} {2024}{\natexlab{b}})}\BibitemShut {NoStop}%
\bibitem [{\citenamefont {Gan}\ \emph {et~al.}(2024)\citenamefont {Gan}, \citenamefont {Fu}, \citenamefont {Wang}, \citenamefont {Xie}, \citenamefont {Gao}, \citenamefont {Wang}, \citenamefont {Chen},\ and\ \citenamefont {Chen}}]{gan2024hydrogenation}%
  \BibitemOpen
  \bibfield  {author} {\bibinfo {author} {\bibfnamefont {G.-R.}\ \bibnamefont {Gan}}, \bibinfo {author} {\bibfnamefont {S.-L.}\ \bibnamefont {Fu}}, \bibinfo {author} {\bibfnamefont {C.-A.}\ \bibnamefont {Wang}}, \bibinfo {author} {\bibfnamefont {Y.-P.}\ \bibnamefont {Xie}}, \bibinfo {author} {\bibfnamefont {X.-L.}\ \bibnamefont {Gao}}, \bibinfo {author} {\bibfnamefont {L.-H.}\ \bibnamefont {Wang}}, \bibinfo {author} {\bibfnamefont {Y.-L.}\ \bibnamefont {Chen}},\ and\ \bibinfo {author} {\bibfnamefont {J.-Y.}\ \bibnamefont {Chen}},\ }\bibfield  {title} {\bibinfo {title} {Hydrogenation-induced superconductivity in monolayer},\ }\href@noop {} {\bibfield  {journal} {\bibinfo  {journal} {EPL}\ }\textbf {\bibinfo {volume} {145}},\ \bibinfo {pages} {56002} (\bibinfo {year} {2024})}\BibitemShut {NoStop}%
\bibitem [{\citenamefont {Fu}\ \emph {et~al.}(2024)\citenamefont {Fu}, \citenamefont {Gan}, \citenamefont {Wang}, \citenamefont {Xie}, \citenamefont {Gao}, \citenamefont {Wang}, \citenamefont {Chen}, \citenamefont {Chen},\ and\ \citenamefont {Wu}}]{fu2024superconductivity}%
  \BibitemOpen
  \bibfield  {author} {\bibinfo {author} {\bibfnamefont {S.-L.}\ \bibnamefont {Fu}}, \bibinfo {author} {\bibfnamefont {G.-R.}\ \bibnamefont {Gan}}, \bibinfo {author} {\bibfnamefont {C.-A.}\ \bibnamefont {Wang}}, \bibinfo {author} {\bibfnamefont {Y.-P.}\ \bibnamefont {Xie}}, \bibinfo {author} {\bibfnamefont {X.-L.}\ \bibnamefont {Gao}}, \bibinfo {author} {\bibfnamefont {L.-H.}\ \bibnamefont {Wang}}, \bibinfo {author} {\bibfnamefont {Y.-L.}\ \bibnamefont {Chen}}, \bibinfo {author} {\bibfnamefont {J.-Y.}\ \bibnamefont {Chen}},\ and\ \bibinfo {author} {\bibfnamefont {X.-Q.}\ \bibnamefont {Wu}},\ }\bibfield  {title} {\bibinfo {title} {Superconductivity in the janus wsh monolayer},\ }\href@noop {} {\bibfield  {journal} {\bibinfo  {journal} {Journal of Superconductivity and Novel Magnetism}\ ,\ \bibinfo {pages} {1}} (\bibinfo {year} {2024})}\BibitemShut {NoStop}%
\bibitem [{\citenamefont {Ul~Hassan}\ and\ \citenamefont {Pinsook}(2024)}]{ul2024superconductivity}%
  \BibitemOpen
  \bibfield  {author} {\bibinfo {author} {\bibfnamefont {M.~M.}\ \bibnamefont {Ul~Hassan}}\ and\ \bibinfo {author} {\bibfnamefont {U.}~\bibnamefont {Pinsook}},\ }\bibfield  {title} {\bibinfo {title} {Superconductivity in monolayer janus titanium-sulfurhydride (tish) at ambient pressure},\ }\href@noop {} {\bibfield  {journal} {\bibinfo  {journal} {Journal of Physics: Condensed Matter}\ } (\bibinfo {year} {2024})}\BibitemShut {NoStop}%
\bibitem [{\citenamefont {Li}\ \emph {et~al.}(2024)\citenamefont {Li}, \citenamefont {Wei}, \citenamefont {Shi}, \citenamefont {Shi}, \citenamefont {Si}, \citenamefont {Liu},\ and\ \citenamefont {Wang}}]{li2024machine}%
  \BibitemOpen
  \bibfield  {author} {\bibinfo {author} {\bibfnamefont {J.}~\bibnamefont {Li}}, \bibinfo {author} {\bibfnamefont {L.}~\bibnamefont {Wei}}, \bibinfo {author} {\bibfnamefont {X.}~\bibnamefont {Shi}}, \bibinfo {author} {\bibfnamefont {L.}~\bibnamefont {Shi}}, \bibinfo {author} {\bibfnamefont {J.}~\bibnamefont {Si}}, \bibinfo {author} {\bibfnamefont {P.-F.}\ \bibnamefont {Liu}},\ and\ \bibinfo {author} {\bibfnamefont {B.-T.}\ \bibnamefont {Wang}},\ }\bibfield  {title} {\bibinfo {title} {Machine learning accelerated discovery of superconducting two-dimensional janus transition metal sulfhydrates},\ }\href@noop {} {\bibfield  {journal} {\bibinfo  {journal} {Physical Review B}\ }\textbf {\bibinfo {volume} {109}},\ \bibinfo {pages} {174516} (\bibinfo {year} {2024})}\BibitemShut {NoStop}%
\bibitem [{\citenamefont {Singh}\ \emph {et~al.}(2021)\citenamefont {Singh}, \citenamefont {Kumbhakar}, \citenamefont {Krishnamoorthy}, \citenamefont {Nakano}, \citenamefont {Sadasivuni}, \citenamefont {Vashishta}, \citenamefont {Roy}, \citenamefont {Kochat},\ and\ \citenamefont {Tiwary}}]{singh2021review}%
  \BibitemOpen
  \bibfield  {author} {\bibinfo {author} {\bibfnamefont {A.~K.}\ \bibnamefont {Singh}}, \bibinfo {author} {\bibfnamefont {P.}~\bibnamefont {Kumbhakar}}, \bibinfo {author} {\bibfnamefont {A.}~\bibnamefont {Krishnamoorthy}}, \bibinfo {author} {\bibfnamefont {A.}~\bibnamefont {Nakano}}, \bibinfo {author} {\bibfnamefont {K.~K.}\ \bibnamefont {Sadasivuni}}, \bibinfo {author} {\bibfnamefont {P.}~\bibnamefont {Vashishta}}, \bibinfo {author} {\bibfnamefont {A.~K.}\ \bibnamefont {Roy}}, \bibinfo {author} {\bibfnamefont {V.}~\bibnamefont {Kochat}},\ and\ \bibinfo {author} {\bibfnamefont {C.~S.}\ \bibnamefont {Tiwary}},\ }\bibfield  {title} {\bibinfo {title} {Review of strategies toward the development of alloy two-dimensional (2d) transition metal dichalcogenides},\ }\href@noop {} {\bibfield  {journal} {\bibinfo  {journal} {Iscience}\ }\textbf {\bibinfo {volume} {24}} (\bibinfo {year} {2021})}\BibitemShut {NoStop}%
\bibitem [{\citenamefont {Chen}\ \emph {et~al.}(2021)\citenamefont {Chen}, \citenamefont {Foo},\ and\ \citenamefont {Tsang}}]{chen2021interstitial}%
  \BibitemOpen
  \bibfield  {author} {\bibinfo {author} {\bibfnamefont {T.}~\bibnamefont {Chen}}, \bibinfo {author} {\bibfnamefont {C.}~\bibnamefont {Foo}},\ and\ \bibinfo {author} {\bibfnamefont {S.~C.~E.}\ \bibnamefont {Tsang}},\ }\bibfield  {title} {\bibinfo {title} {Interstitial and substitutional light elements in transition metals for heterogeneous catalysis},\ }\href@noop {} {\bibfield  {journal} {\bibinfo  {journal} {Chemical science}\ }\textbf {\bibinfo {volume} {12}},\ \bibinfo {pages} {517} (\bibinfo {year} {2021})}\BibitemShut {NoStop}%
\bibitem [{\citenamefont {Xie}\ \emph {et~al.}(2024)\citenamefont {Xie}, \citenamefont {Huang}, \citenamefont {Zhao}, \citenamefont {Huang}, \citenamefont {Li}, \citenamefont {Gu},\ and\ \citenamefont {Zeng}}]{xie2024strong}%
  \BibitemOpen
  \bibfield  {author} {\bibinfo {author} {\bibfnamefont {H.}~\bibnamefont {Xie}}, \bibinfo {author} {\bibfnamefont {Z.}~\bibnamefont {Huang}}, \bibinfo {author} {\bibfnamefont {Y.}~\bibnamefont {Zhao}}, \bibinfo {author} {\bibfnamefont {H.}~\bibnamefont {Huang}}, \bibinfo {author} {\bibfnamefont {G.}~\bibnamefont {Li}}, \bibinfo {author} {\bibfnamefont {Z.}~\bibnamefont {Gu}},\ and\ \bibinfo {author} {\bibfnamefont {S.}~\bibnamefont {Zeng}},\ }\bibfield  {title} {\bibinfo {title} {Strong electron--phonon coupling and multigap superconductivity in 2h/1t janus mosli monolayer},\ }\href@noop {} {\bibfield  {journal} {\bibinfo  {journal} {The Journal of Chemical Physics}\ }\textbf {\bibinfo {volume} {160}} (\bibinfo {year} {2024})}\BibitemShut {NoStop}%
\bibitem [{\citenamefont {Bardeen}\ \emph {et~al.}(1957)\citenamefont {Bardeen}, \citenamefont {Cooper},\ and\ \citenamefont {Schrieffer}}]{bardeen1957microscopic}%
  \BibitemOpen
  \bibfield  {author} {\bibinfo {author} {\bibfnamefont {J.}~\bibnamefont {Bardeen}}, \bibinfo {author} {\bibfnamefont {L.~N.}\ \bibnamefont {Cooper}},\ and\ \bibinfo {author} {\bibfnamefont {J.~R.}\ \bibnamefont {Schrieffer}},\ }\bibfield  {title} {\bibinfo {title} {Microscopic theory of superconductivity},\ }\href@noop {} {\bibfield  {journal} {\bibinfo  {journal} {Physical Review}\ }\textbf {\bibinfo {volume} {106}},\ \bibinfo {pages} {162} (\bibinfo {year} {1957})}\BibitemShut {NoStop}%
\bibitem [{\citenamefont {Frohlich}(1950)}]{frohlich1950theory}%
  \BibitemOpen
  \bibfield  {author} {\bibinfo {author} {\bibfnamefont {H.}~\bibnamefont {Frohlich}},\ }\bibfield  {title} {\bibinfo {title} {Theory of the superconducting state. i. the ground state at the absolute zero of temperature},\ }\href@noop {} {\bibfield  {journal} {\bibinfo  {journal} {Physical Review}\ }\textbf {\bibinfo {volume} {79}},\ \bibinfo {pages} {845} (\bibinfo {year} {1950})}\BibitemShut {NoStop}%
\bibitem [{\citenamefont {Migdal}(1958)}]{migdal1958interaction}%
  \BibitemOpen
  \bibfield  {author} {\bibinfo {author} {\bibfnamefont {A.}~\bibnamefont {Migdal}},\ }\bibfield  {title} {\bibinfo {title} {Interaction between electrons and lattice vibrations in a normal metal},\ }\href@noop {} {\bibfield  {journal} {\bibinfo  {journal} {Sov. Phys. JETP}\ }\textbf {\bibinfo {volume} {7}},\ \bibinfo {pages} {996} (\bibinfo {year} {1958})}\BibitemShut {NoStop}%
\bibitem [{\citenamefont {Eliashberg}(1960)}]{eliashberg1960interactions}%
  \BibitemOpen
  \bibfield  {author} {\bibinfo {author} {\bibfnamefont {G.}~\bibnamefont {Eliashberg}},\ }\bibfield  {title} {\bibinfo {title} {Interactions between electrons and lattice vibrations in a superconductor},\ }\href@noop {} {\bibfield  {journal} {\bibinfo  {journal} {Sov. Phys. JETP}\ }\textbf {\bibinfo {volume} {11}},\ \bibinfo {pages} {696} (\bibinfo {year} {1960})}\BibitemShut {NoStop}%
\bibitem [{\citenamefont {Nambu}(1960)}]{nambu1960quasi}%
  \BibitemOpen
  \bibfield  {author} {\bibinfo {author} {\bibfnamefont {Y.}~\bibnamefont {Nambu}},\ }\bibfield  {title} {\bibinfo {title} {Quasi-particles and gauge invariance in the theory of superconductivity},\ }\href@noop {} {\bibfield  {journal} {\bibinfo  {journal} {Physical Review}\ }\textbf {\bibinfo {volume} {117}},\ \bibinfo {pages} {648} (\bibinfo {year} {1960})}\BibitemShut {NoStop}%
\bibitem [{\citenamefont {Giannozzi}\ \emph {et~al.}(2009)\citenamefont {Giannozzi}, \citenamefont {Baroni}, \citenamefont {Bonini}, \citenamefont {Calandra}, \citenamefont {Car}, \citenamefont {Cavazzoni}, \citenamefont {Ceresoli}, \citenamefont {Chiarotti}, \citenamefont {Cococcioni}, \citenamefont {Dabo} \emph {et~al.}}]{giannozzi2009quantum}%
  \BibitemOpen
  \bibfield  {author} {\bibinfo {author} {\bibfnamefont {P.}~\bibnamefont {Giannozzi}}, \bibinfo {author} {\bibfnamefont {S.}~\bibnamefont {Baroni}}, \bibinfo {author} {\bibfnamefont {N.}~\bibnamefont {Bonini}}, \bibinfo {author} {\bibfnamefont {M.}~\bibnamefont {Calandra}}, \bibinfo {author} {\bibfnamefont {R.}~\bibnamefont {Car}}, \bibinfo {author} {\bibfnamefont {C.}~\bibnamefont {Cavazzoni}}, \bibinfo {author} {\bibfnamefont {D.}~\bibnamefont {Ceresoli}}, \bibinfo {author} {\bibfnamefont {G.~L.}\ \bibnamefont {Chiarotti}}, \bibinfo {author} {\bibfnamefont {M.}~\bibnamefont {Cococcioni}}, \bibinfo {author} {\bibfnamefont {I.}~\bibnamefont {Dabo}}, \emph {et~al.},\ }\bibfield  {title} {\bibinfo {title} {Quantum espresso: a modular and open-source software project for quantum simulations of materials},\ }\href@noop {} {\bibfield  {journal} {\bibinfo  {journal} {Journal of physics: Condensed matter}\ }\textbf {\bibinfo {volume} {21}},\ \bibinfo {pages} {395502} (\bibinfo {year} {2009})}\BibitemShut {NoStop}%
\bibitem [{\citenamefont {Giannozzi}\ \emph {et~al.}(2017)\citenamefont {Giannozzi}, \citenamefont {Andreussi}, \citenamefont {Brumme}, \citenamefont {Bunau}, \citenamefont {Nardelli}, \citenamefont {Calandra}, \citenamefont {Car}, \citenamefont {Cavazzoni}, \citenamefont {Ceresoli}, \citenamefont {Cococcioni} \emph {et~al.}}]{giannozzi2017advanced}%
  \BibitemOpen
  \bibfield  {author} {\bibinfo {author} {\bibfnamefont {P.}~\bibnamefont {Giannozzi}}, \bibinfo {author} {\bibfnamefont {O.}~\bibnamefont {Andreussi}}, \bibinfo {author} {\bibfnamefont {T.}~\bibnamefont {Brumme}}, \bibinfo {author} {\bibfnamefont {O.}~\bibnamefont {Bunau}}, \bibinfo {author} {\bibfnamefont {M.~B.}\ \bibnamefont {Nardelli}}, \bibinfo {author} {\bibfnamefont {M.}~\bibnamefont {Calandra}}, \bibinfo {author} {\bibfnamefont {R.}~\bibnamefont {Car}}, \bibinfo {author} {\bibfnamefont {C.}~\bibnamefont {Cavazzoni}}, \bibinfo {author} {\bibfnamefont {D.}~\bibnamefont {Ceresoli}}, \bibinfo {author} {\bibfnamefont {M.}~\bibnamefont {Cococcioni}}, \emph {et~al.},\ }\bibfield  {title} {\bibinfo {title} {Advanced capabilities for materials modelling with quantum espresso},\ }\href@noop {} {\bibfield  {journal} {\bibinfo  {journal} {Journal of physics: Condensed matter}\ }\textbf {\bibinfo {volume} {29}},\ \bibinfo {pages} {465901} (\bibinfo {year} {2017})}\BibitemShut {NoStop}%
\bibitem [{\citenamefont {Giustino}(2017)}]{giustino2017electron}%
  \BibitemOpen
  \bibfield  {author} {\bibinfo {author} {\bibfnamefont {F.}~\bibnamefont {Giustino}},\ }\bibfield  {title} {\bibinfo {title} {Electron-phonon interactions from first principles},\ }\href@noop {} {\bibfield  {journal} {\bibinfo  {journal} {Reviews of Modern Physics}\ }\textbf {\bibinfo {volume} {89}},\ \bibinfo {pages} {015003} (\bibinfo {year} {2017})}\BibitemShut {NoStop}%
\bibitem [{\citenamefont {Giustino}\ \emph {et~al.}(2007)\citenamefont {Giustino}, \citenamefont {Cohen},\ and\ \citenamefont {Louie}}]{giustino2007electron}%
  \BibitemOpen
  \bibfield  {author} {\bibinfo {author} {\bibfnamefont {F.}~\bibnamefont {Giustino}}, \bibinfo {author} {\bibfnamefont {M.~L.}\ \bibnamefont {Cohen}},\ and\ \bibinfo {author} {\bibfnamefont {S.~G.}\ \bibnamefont {Louie}},\ }\bibfield  {title} {\bibinfo {title} {Electron-phonon interaction using wannier functions},\ }\href@noop {} {\bibfield  {journal} {\bibinfo  {journal} {Physical Review B}\ }\textbf {\bibinfo {volume} {76}},\ \bibinfo {pages} {165108} (\bibinfo {year} {2007})}\BibitemShut {NoStop}%
\bibitem [{\citenamefont {Noffsinger}\ \emph {et~al.}(2010)\citenamefont {Noffsinger}, \citenamefont {Giustino}, \citenamefont {Malone}, \citenamefont {Park}, \citenamefont {Louie},\ and\ \citenamefont {Cohen}}]{noffsinger2010epw}%
  \BibitemOpen
  \bibfield  {author} {\bibinfo {author} {\bibfnamefont {J.}~\bibnamefont {Noffsinger}}, \bibinfo {author} {\bibfnamefont {F.}~\bibnamefont {Giustino}}, \bibinfo {author} {\bibfnamefont {B.~D.}\ \bibnamefont {Malone}}, \bibinfo {author} {\bibfnamefont {C.-H.}\ \bibnamefont {Park}}, \bibinfo {author} {\bibfnamefont {S.~G.}\ \bibnamefont {Louie}},\ and\ \bibinfo {author} {\bibfnamefont {M.~L.}\ \bibnamefont {Cohen}},\ }\bibfield  {title} {\bibinfo {title} {Epw: A program for calculating the electron--phonon coupling using maximally localized wannier functions},\ }\href@noop {} {\bibfield  {journal} {\bibinfo  {journal} {Computer Physics Communications}\ }\textbf {\bibinfo {volume} {181}},\ \bibinfo {pages} {2140} (\bibinfo {year} {2010})}\BibitemShut {NoStop}%
\bibitem [{\citenamefont {Ponce}\ \emph {et~al.}(2016)\citenamefont {Ponce}, \citenamefont {Margine}, \citenamefont {Verdi},\ and\ \citenamefont {Giustino}}]{ponce2016epw}%
  \BibitemOpen
  \bibfield  {author} {\bibinfo {author} {\bibfnamefont {S.}~\bibnamefont {Ponce}}, \bibinfo {author} {\bibfnamefont {E.~R.}\ \bibnamefont {Margine}}, \bibinfo {author} {\bibfnamefont {C.}~\bibnamefont {Verdi}},\ and\ \bibinfo {author} {\bibfnamefont {F.}~\bibnamefont {Giustino}},\ }\bibfield  {title} {\bibinfo {title} {Epw: Electron--phonon coupling, transport and superconducting properties using maximally localized wannier functions},\ }\href@noop {} {\bibfield  {journal} {\bibinfo  {journal} {Computer Physics Communications}\ }\textbf {\bibinfo {volume} {209}},\ \bibinfo {pages} {116} (\bibinfo {year} {2016})}\BibitemShut {NoStop}%
\bibitem [{\citenamefont {Momma}\ and\ \citenamefont {Izumi}(2011)}]{momma2011vesta}%
  \BibitemOpen
  \bibfield  {author} {\bibinfo {author} {\bibfnamefont {K.}~\bibnamefont {Momma}}\ and\ \bibinfo {author} {\bibfnamefont {F.}~\bibnamefont {Izumi}},\ }\bibfield  {title} {\bibinfo {title} {Vesta 3 for three-dimensional visualization of crystal, volumetric and morphology data},\ }\href@noop {} {\bibfield  {journal} {\bibinfo  {journal} {Journal of applied crystallography}\ }\textbf {\bibinfo {volume} {44}},\ \bibinfo {pages} {1272} (\bibinfo {year} {2011})}\BibitemShut {NoStop}%
\bibitem [{\citenamefont {Pfrommer}\ \emph {et~al.}(1997)\citenamefont {Pfrommer}, \citenamefont {Cote}, \citenamefont {Louie},\ and\ \citenamefont {Cohen}}]{BFGS}%
  \BibitemOpen
  \bibfield  {author} {\bibinfo {author} {\bibfnamefont {B.~G.}\ \bibnamefont {Pfrommer}}, \bibinfo {author} {\bibfnamefont {M.}~\bibnamefont {Cote}}, \bibinfo {author} {\bibfnamefont {S.~G.}\ \bibnamefont {Louie}},\ and\ \bibinfo {author} {\bibfnamefont {M.~L.}\ \bibnamefont {Cohen}},\ }\bibfield  {title} {\bibinfo {title} {Relaxation of crystals with the quasi-{N}ewton method},\ }\href@noop {} {\bibfield  {journal} {\bibinfo  {journal} {J. Comput. Phys.}\ }\textbf {\bibinfo {volume} {131}},\ \bibinfo {pages} {233} (\bibinfo {year} {1997})}\BibitemShut {NoStop}%
\bibitem [{\citenamefont {Liu}\ and\ \citenamefont {Nocedal}(1989)}]{liu1989limited}%
  \BibitemOpen
  \bibfield  {author} {\bibinfo {author} {\bibfnamefont {D.~C.}\ \bibnamefont {Liu}}\ and\ \bibinfo {author} {\bibfnamefont {J.}~\bibnamefont {Nocedal}},\ }\bibfield  {title} {\bibinfo {title} {On the limited memory bfgs method for large scale optimization},\ }\href@noop {} {\bibfield  {journal} {\bibinfo  {journal} {Mathematical programming}\ }\textbf {\bibinfo {volume} {45}},\ \bibinfo {pages} {503} (\bibinfo {year} {1989})}\BibitemShut {NoStop}%
\bibitem [{\citenamefont {Hamann}(2013)}]{hamann2013optimized}%
  \BibitemOpen
  \bibfield  {author} {\bibinfo {author} {\bibfnamefont {D.}~\bibnamefont {Hamann}},\ }\bibfield  {title} {\bibinfo {title} {Optimized norm-conserving vanderbilt pseudopotentials},\ }\href@noop {} {\bibfield  {journal} {\bibinfo  {journal} {Physical Review B}\ }\textbf {\bibinfo {volume} {88}},\ \bibinfo {pages} {085117} (\bibinfo {year} {2013})}\BibitemShut {NoStop}%
\bibitem [{\citenamefont {Schlipf}\ and\ \citenamefont {Gygi}(2015)}]{schlipf2015optimization}%
  \BibitemOpen
  \bibfield  {author} {\bibinfo {author} {\bibfnamefont {M.}~\bibnamefont {Schlipf}}\ and\ \bibinfo {author} {\bibfnamefont {F.}~\bibnamefont {Gygi}},\ }\bibfield  {title} {\bibinfo {title} {Optimization algorithm for the generation of oncv pseudopotentials},\ }\href@noop {} {\bibfield  {journal} {\bibinfo  {journal} {Computer Physics Communications}\ }\textbf {\bibinfo {volume} {196}},\ \bibinfo {pages} {36} (\bibinfo {year} {2015})}\BibitemShut {NoStop}%
\bibitem [{\citenamefont {Perdew}\ \emph {et~al.}(1996)\citenamefont {Perdew}, \citenamefont {Burke},\ and\ \citenamefont {Ernzerhof}}]{perdew1996generalized}%
  \BibitemOpen
  \bibfield  {author} {\bibinfo {author} {\bibfnamefont {J.~P.}\ \bibnamefont {Perdew}}, \bibinfo {author} {\bibfnamefont {K.}~\bibnamefont {Burke}},\ and\ \bibinfo {author} {\bibfnamefont {M.}~\bibnamefont {Ernzerhof}},\ }\bibfield  {title} {\bibinfo {title} {Generalized gradient approximation made simple},\ }\href@noop {} {\bibfield  {journal} {\bibinfo  {journal} {Physical review letters}\ }\textbf {\bibinfo {volume} {77}},\ \bibinfo {pages} {3865} (\bibinfo {year} {1996})}\BibitemShut {NoStop}%
\bibitem [{\citenamefont {Monkhorst}\ and\ \citenamefont {Pack}(1976)}]{monkhorst1976special}%
  \BibitemOpen
  \bibfield  {author} {\bibinfo {author} {\bibfnamefont {H.~J.}\ \bibnamefont {Monkhorst}}\ and\ \bibinfo {author} {\bibfnamefont {J.~D.}\ \bibnamefont {Pack}},\ }\bibfield  {title} {\bibinfo {title} {Special points for brillouin-zone integrations},\ }\href@noop {} {\bibfield  {journal} {\bibinfo  {journal} {Physical review B}\ }\textbf {\bibinfo {volume} {13}},\ \bibinfo {pages} {5188} (\bibinfo {year} {1976})}\BibitemShut {NoStop}%
\bibitem [{\citenamefont {Methfessel}\ and\ \citenamefont {Paxton}(1989)}]{methfessel1989high}%
  \BibitemOpen
  \bibfield  {author} {\bibinfo {author} {\bibfnamefont {M.}~\bibnamefont {Methfessel}}\ and\ \bibinfo {author} {\bibfnamefont {A.}~\bibnamefont {Paxton}},\ }\bibfield  {title} {\bibinfo {title} {High-precision sampling for brillouin-zone integration in metals},\ }\href@noop {} {\bibfield  {journal} {\bibinfo  {journal} {physical review B}\ }\textbf {\bibinfo {volume} {40}},\ \bibinfo {pages} {3616} (\bibinfo {year} {1989})}\BibitemShut {NoStop}%
\bibitem [{\citenamefont {Kokalj}(2003)}]{kokalj2003computer}%
  \BibitemOpen
  \bibfield  {author} {\bibinfo {author} {\bibfnamefont {A.}~\bibnamefont {Kokalj}},\ }\bibfield  {title} {\bibinfo {title} {Computer graphics and graphical user interfaces as tools in simulations of matter at the atomic scale},\ }\href@noop {} {\bibfield  {journal} {\bibinfo  {journal} {Computational Materials Science}\ }\textbf {\bibinfo {volume} {28}},\ \bibinfo {pages} {155} (\bibinfo {year} {2003})}\BibitemShut {NoStop}%
\bibitem [{\citenamefont {Margine}\ and\ \citenamefont {Giustino}(2013)}]{margine2013anisotropic}%
  \BibitemOpen
  \bibfield  {author} {\bibinfo {author} {\bibfnamefont {E.~R.}\ \bibnamefont {Margine}}\ and\ \bibinfo {author} {\bibfnamefont {F.}~\bibnamefont {Giustino}},\ }\bibfield  {title} {\bibinfo {title} {Anisotropic migdal-eliashberg theory using wannier functions},\ }\href@noop {} {\bibfield  {journal} {\bibinfo  {journal} {Physical Review B}\ }\textbf {\bibinfo {volume} {87}},\ \bibinfo {pages} {024505} (\bibinfo {year} {2013})}\BibitemShut {NoStop}%
\bibitem [{\citenamefont {Allen}\ and\ \citenamefont {Dynes}(1975)}]{allen1975transition}%
  \BibitemOpen
  \bibfield  {author} {\bibinfo {author} {\bibfnamefont {P.~B.}\ \bibnamefont {Allen}}\ and\ \bibinfo {author} {\bibfnamefont {R.}~\bibnamefont {Dynes}},\ }\bibfield  {title} {\bibinfo {title} {Transition temperature of strong-coupled superconductors reanalyzed},\ }\href@noop {} {\bibfield  {journal} {\bibinfo  {journal} {Physical Review B}\ }\textbf {\bibinfo {volume} {12}},\ \bibinfo {pages} {905} (\bibinfo {year} {1975})}\BibitemShut {NoStop}%
\bibitem [{\citenamefont {Margine}\ and\ \citenamefont {Giustino}(2014)}]{margine2014two}%
  \BibitemOpen
  \bibfield  {author} {\bibinfo {author} {\bibfnamefont {E.}~\bibnamefont {Margine}}\ and\ \bibinfo {author} {\bibfnamefont {F.}~\bibnamefont {Giustino}},\ }\bibfield  {title} {\bibinfo {title} {Two-gap superconductivity in heavily n-doped graphene: Ab initio migdal-eliashberg theory},\ }\href@noop {} {\bibfield  {journal} {\bibinfo  {journal} {Physical Review B}\ }\textbf {\bibinfo {volume} {90}},\ \bibinfo {pages} {014518} (\bibinfo {year} {2014})}\BibitemShut {NoStop}%
\bibitem [{\citenamefont {Zhao}\ \emph {et~al.}(2019)\citenamefont {Zhao}, \citenamefont {Lian}, \citenamefont {Zeng}, \citenamefont {Dai}, \citenamefont {Meng},\ and\ \citenamefont {Ni}}]{zhao2019two}%
  \BibitemOpen
  \bibfield  {author} {\bibinfo {author} {\bibfnamefont {Y.}~\bibnamefont {Zhao}}, \bibinfo {author} {\bibfnamefont {C.}~\bibnamefont {Lian}}, \bibinfo {author} {\bibfnamefont {S.}~\bibnamefont {Zeng}}, \bibinfo {author} {\bibfnamefont {Z.}~\bibnamefont {Dai}}, \bibinfo {author} {\bibfnamefont {S.}~\bibnamefont {Meng}},\ and\ \bibinfo {author} {\bibfnamefont {J.}~\bibnamefont {Ni}},\ }\bibfield  {title} {\bibinfo {title} {Two-gap and three-gap superconductivity in alb 2-based films},\ }\href@noop {} {\bibfield  {journal} {\bibinfo  {journal} {Physical Review B}\ }\textbf {\bibinfo {volume} {100}},\ \bibinfo {pages} {094516} (\bibinfo {year} {2019})}\BibitemShut {NoStop}%
\bibitem [{\citenamefont {Gao}\ \emph {et~al.}(2020)\citenamefont {Gao}, \citenamefont {Yan}, \citenamefont {Lu},\ and\ \citenamefont {Xiang}}]{gao2020strong}%
  \BibitemOpen
  \bibfield  {author} {\bibinfo {author} {\bibfnamefont {M.}~\bibnamefont {Gao}}, \bibinfo {author} {\bibfnamefont {X.-W.}\ \bibnamefont {Yan}}, \bibinfo {author} {\bibfnamefont {Z.-Y.}\ \bibnamefont {Lu}},\ and\ \bibinfo {author} {\bibfnamefont {T.}~\bibnamefont {Xiang}},\ }\bibfield  {title} {\bibinfo {title} {Strong-coupling superconductivity in lib 2 c 2 trilayer films},\ }\href@noop {} {\bibfield  {journal} {\bibinfo  {journal} {Physical Review B}\ }\textbf {\bibinfo {volume} {101}},\ \bibinfo {pages} {094501} (\bibinfo {year} {2020})}\BibitemShut {NoStop}%
\bibitem [{\citenamefont {Modak}\ \emph {et~al.}(2021)\citenamefont {Modak}, \citenamefont {Verma},\ and\ \citenamefont {Mishra}}]{modak2021prediction}%
  \BibitemOpen
  \bibfield  {author} {\bibinfo {author} {\bibfnamefont {P.}~\bibnamefont {Modak}}, \bibinfo {author} {\bibfnamefont {A.~K.}\ \bibnamefont {Verma}},\ and\ \bibinfo {author} {\bibfnamefont {A.~K.}\ \bibnamefont {Mishra}},\ }\bibfield  {title} {\bibinfo {title} {Prediction of superconductivity at 70 k in a pristine monolayer of libc},\ }\href@noop {} {\bibfield  {journal} {\bibinfo  {journal} {Physical Review B}\ }\textbf {\bibinfo {volume} {104}},\ \bibinfo {pages} {054504} (\bibinfo {year} {2021})}\BibitemShut {NoStop}%
\end{thebibliography}%

\end{document}